\documentclass[apj]{emulateapj}
\usepackage{graphicx}
\usepackage{amssymb}
\usepackage{amsmath}
\usepackage{epstopdf}
\usepackage{color}

\shorttitle{Viscous sloshing CFs in Virgo}
\shortauthors{Roediger et al.~2012}

\begin{document}

\newcommand{\degree}{^o}
\newcommand{\subsun}{_{\sun}}
\newcommand{\KHI}{_{\mathrm{KHI}}}
\newcommand{\CF}{_{\mathrm{CF}}}
\newcommand{\Hot}{_{\mathrm{hot}}}
\newcommand{\Cold}{_{\mathrm{cold}}}
\newcommand{\Max}{_{\mathrm{max}}}
\newcommand{\ICM}{_{\mathrm{ICM}}}
\newcommand{\Cross}{_{\mathrm{cross}}}
\newcommand{\KeV}{\,\textrm{keV}}
\newcommand{\Kpc}{\,\textrm{kpc}}
\newcommand{\Kms}{\,\textrm{km}\,\textrm{s}^{-1}}
\newcommand{\K}{\,\mathrm{K}}
\newcommand{\ccm}{\,\mathrm{cm}^{-3}}
\newcommand{\gccm}{\,\mathrm{g}\,\mathrm{cm}^{-3}}
\newcommand{\cmss}{\,\mathrm{cm}\,\mathrm{s}^{-2}}
\newcommand{\Myr}{\,\mathrm{Myr}}
\newcommand{\Gyr}{\,\mathrm{Gyr}}
\newcommand{\Reyn}{\textrm{Re}}

\definecolor{rred}{rgb}{1,0,0}
\definecolor{bblue}{rgb}{0,0,1}

\title{Kelvin-Helmholtz instabilities at the sloshing cold fronts in the Virgo cluster as a measure for the effective ICM viscosity}
\author{E.~Roediger\altaffilmark{1,2,3}, R.~P.~Kraft\altaffilmark{3}, W.~R.~Forman\altaffilmark{3}, P.~E.~J.~Nulsen\altaffilmark{3}, E.~Churazov\altaffilmark{4}}
\affil{
\altaffilmark{1}Hamburger Sternwarte, Universit\"at Hamburg, Gojensbergsweg 112, D-21029 Hamburg, Germany \newline
\altaffilmark{2}Jacobs University Bremen, Campus Ring 1, 28759 Bremen, Germany\newline
\altaffilmark{3}Harvard/Smithsonian Center for Astrophysics, 60 Garden Street, Cambridge, MA 02138, USA\newline
\altaffilmark{4}Max-Planck-Institut f\"ur Astrophysik, Karl-Schwarzschild-Str. 1, 85748 Garching, Germany
}
\email{eroediger@hs.uni-hamburg.de}

\begin{abstract}
Sloshing cold fronts (CFs) arise from minor merger triggered gas sloshing. Their detailed structure depends on the properties of the intra-cluster medium (ICM): 
hydrodynamical simulations predict the CFs to be distorted by Kelvin-Helmholtz instabilities (KHIs), but aligned magnetic fields,  viscosity, or thermal conduction can suppress the KHIs. Thus, observing the detailed structure of sloshing CFs can be used to constrain these ICM properties. Both smooth and distorted sloshing CFs have been observed, indicating that the KHI  is suppressed in some clusters, but not in all. Consequently, we need to address at least some sloshing clusters individually before drawing general conclusions about the ICM properties.
We present the first detailed attempt to constrain the ICM properties in a specific cluster from the structure of its sloshing CF. Proximity and brightness make the Virgo cluster an ideal target. We combine observations and Virgo-specific hydrodynamical sloshing simulations.  Here we focus on a Spitzer-like temperature dependent viscosity as a mechanism to suppress the KHI, but discuss the alternative mechanisms in detail. 
We identify the CF at 90$\Kpc$ north and north-east  of the Virgo center as the best location in the cluster to observe a possible KHI suppression. For viscosities  $\gtrsim 10\%$ of the Spitzer value KHIs at this CF are suppressed. We describe in detail the observable signatures at low and high viscosities, i.e.~in the presence or absence of KHIs. We find indications for a low ICM viscosity in archival \textit{XMM-Newton} data and demonstrate the detectability of the predicted features in deep \textit{Chandra} observations. 
\end{abstract}

\maketitle

\section{Introduction} \label{sec:intro}
%
Viscous stresses and thermal conduction in the intra-cluster medium (ICM) of clusters of galaxies play a critical role in shaping the thermal structure of the cluster gas, in dissipating and redistributing  the  energy released by mergers and AGN outbursts, and controlling the feedback between the ICM and the AGN inside the Bondi radius.  In general, the transport processes in the gas must play a key role in the formation of structure down to scales of galaxies and star formation.

The magnitude of the effective transport coefficients however is one of the longstanding and still unresolved questions. In an unmagnetized plasma, under the conditions of the ICM, both thermal conductivity and viscosity are large due to the strong temperature dependence $\propto T^{5/2}$ (\citealt{Spitzer1956,Sarazin1988}). For example,~typical Reynolds numbers in the ICM are
\begin{eqnarray}
\Reyn & =& 7 \; f_{\mu}^{-1}\; \left(\frac{U}{300\Kms}\right)\; \left(\frac{\lambda}{10\Kpc}\right) \times \nonumber\\
&& \left(\frac{n\ICM}{3\times 10^{-3}\ccm}\right) \; \left( \frac{kT\ICM}{3\KeV}  \right)^{-5/2},
\label{eq:Re}
\end{eqnarray}
for length scales of $\lambda$, velocities $U$, particle density $n\ICM$ and ICM temperature $T\ICM$. In this equation we have included the suppression factor $f_{\mu}$ to parametrize an unknown degree of suppression of the viscosity. For a full Spitzer viscosity we have $f_{\mu}=1$. However, even weak  magnetic fields strongly suppress diffusion of heat and momentum perpendicular to the field, and tangled magnetic fields may lead to an overall suppression of conduction and viscosity. The degree of suppression depends on the magnetic field structure and could be moderate ($f_{\mu}\sim 0.3$) or significant with $f_{\mu}\ll 1$ (\citealt{Narayan2001}). The magnitude of the effective transport coefficients is relevant for numerous processes in galaxy clusters, e.g., the morphology and evolution of AGN inflated cavities (\citealt{Reynolds2005,Kaiser2005}), the dissipation of AGN induced sound waves and thus balancing radiative cooling in cool cores  (\citealt{Fabian2005}), and gas stripping from galaxies (\citealt{Nulsen1982}). 
On scales smaller than the coherence length, diffusion of energy and momentum should be anisotropic, preferentially along the field lines. This can in itself lead to interesting effects and new instabilities like the magneto-thermal instability and the heat flux driven buoyancy instability (e.g., \citealt{Balbus2010a,Parrish2010,Kunz2011,Kunz2012}). The relevance of these processes for different aspects of ICM physics is a topic currently being investigated by different groups.

There are many theoretical arguments regarding the magnitude of the effective ICM viscosity, but the observational evidence in the cluster ICM is controversial.  For example, the coherent morphology of AGN cavities can be explained by a substantial ICM viscosity (\citealt{Reynolds2005,Fabian2003halpha}), but turbulence in cluster cores has been invoked as a mechanism to amplify tiny primordial magnetic fields to the  values observed today (\citealt{Ryu2011}), and to broaden the heavy element abundance peaks in cluster centers (\citealt{Rebusco2005,Rebusco2006,Roediger2007bubbles}). Several observations have been made to search for turbulence in the ICM.  X-ray studies of the ICM in the Coma cluster estimated the power spectrum of density fluctuations in the gas by measuring the spatial variation of the surface brightness fluctuations \citep{Schuecker2004, Churazov2012}. Turbulence in the ICM is one possible origin of these fluctuations.  Additionally, attempts to measure the turbulent velocity broadening of emission lines in clusters and early-type galaxies have been made with the \textit{XMM-Newton} RGS \citep{Sanders2010a}.  Future high spectral resolution observations will allow us to constrain the velocity power spectrum of the ICM by measuring the centroids and widths of X-ray emission lines \citep{Zhuravleva2012}.  

\begin{figure}
\includegraphics[width=0.49\textwidth]{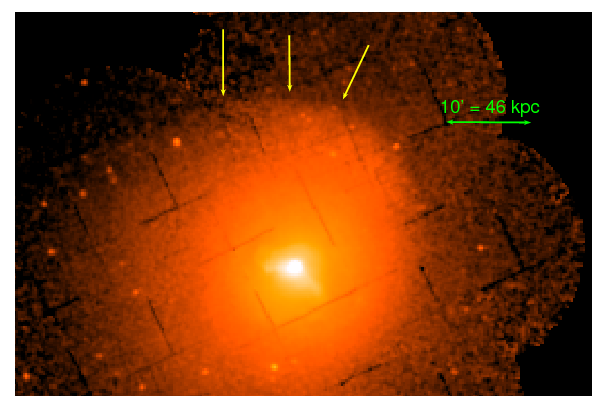}
\caption{\textit{XMM-Newton} MOS1+2 image of the ICM of the Virgo cluster, background subtracted and exposure corrected (0.5-3.0 keV band).  M87 lies at the brightness peak. The sloshing cold front to the north (N) of the nucleus is denoted by the yellow arrows. It clearly continues to the north-west (NW) and somewhat more diffuse to the north-east (NE).}
\label{fig:virgocf}
\end{figure}

There are some places in galaxy clusters where we may directly witness the impact of the effective viscosity, cold fronts being one of them. Sloshing cold fronts (CFs) are density and temperature discontinuities in the ICM of galaxy clusters (see review by \citealt{Markevitch2007}, also for how they are distinguished from merger CFs). They are observed as arc-like edges in surface brightness, temperature and metallicity maps, wrapped around the cluster cores in spiral-like or staggered arc-like patterns (e.g., Figs.~\ref{fig:virgocf} and \ref{fig:Xray} later in this paper). They are attributed to sloshing or oscillatory motions of the ICM inside the cluster potentials, triggered by a  merger that was too weak to destroy the cluster core (\citealt{Markevitch2001,Ascasibar2006}). Such sloshing  CFs are commonly seen in clusters (\citealt{Markevitch2003,Ghizzardi2010}) and  have been observed in detail in several clusters and groups (see \citealt{Markevitch2007}, \citealt{Roediger2012a496} for lists of examples, also Sect.~\ref{sec:compare_others}). 

Sloshing CFs are accompanied by shear flows along them and thus should be subject to the Kelvin-Helmholtz instability (KHI), as confirmed by high resolution hydrodynamical simulations (\citealt{ZuHone2010,Roediger2011,Roediger2012a496}; or Fig.~\ref{fig:tempslices} later in this paper). The growth of KHI modes above a critical wavelength  is suppressed by gravity (\citealt{Chandrasekhar1961}). The presence of $\sim 10$ to 50 kpc KHI modes in the hydrodynamical simulations demonstrates that this critical wavelength is sufficiently large to allow the growth of KHIs at observable length scales (the same can be derived analytically).  Furthermore, KHIs can be suppressed at all wavelengths by sufficiently strong magnetic fields aligned with the fronts (\citealt{Chandrasekhar1961,Vikhlinin2002,Keshet2010,ZuHone2011}), or at small wavelengths by a sufficiently strong viscosity. Consequently, the presence or absence of KHIs at sloshing CFs can be used to constrain these ICM properties. Examples of both smooth CFs and distorted ones, resembling the KHI, are known (see  Sect.~\ref{sec:compare_others} for examples). 
While \citet{ZuHone2010,ZuHone2011,ZuHone2012} have studied the global  impact of viscosity, magnetic fields and thermal conduction on the sloshing process, a detailed attempt to derive the ICM properties from the observed structure of sloshing CFs has not been performed yet. This is the aim of this and a series of forthcoming papers.

At a distance of only 17 Mpc (1$''$=80 pc) the Virgo cluster is the nearest galaxy cluster and can thus be observed with high resolution. The \textit{XMM-Newton} mosaic around M87 in the center of the Virgo cluster  displays a sharp edge in surface brightness  at 90 kpc north-west (NW) of the cluster core as shown in Fig.~\ref{fig:virgocf}. Using a subset of this data, \citet{Simionescu2010} demonstrated that this feature is a sloshing CF. A detailed analysis revealed a secondary and tertiary CF at 30 kpc and 20 kpc to the south-east and NW, respectively.  \citet{Roediger2011} simulated ICM sloshing specifically for the Virgo cluster and inferred the most likely  merger scenario from a detailed comparison to the observations. The proximity of Virgo, the high X-ray surface brightness and favorable merger geometry in the plane of the sky allow us to probe the shear flow interface with a spatial resolution of $\sim$100 pc. 

We have re-simulated the gas sloshing in Virgo at high resolution, with different levels of ICM viscosity as described in Sect.~\ref{sec:method}. We demonstrate that even a viscosity of about $10\%$ of the Spitzer value efficiently suppresses the KHI in Sect.~\ref{sec:khsuppr}, and compare the simulation results to an analytical estimate. Sect.~\ref{sec:observable} describes  observable features at low and high viscosity. We report marginal evidence for a low ICM viscosity in the existing shallow \textit{XMM-Newton} observation (Section~\ref{sec:evidence}) of the Virgo cluster CF.  We demonstrate that the presence or absence of KHIs can be detected with deep  \textit{Chandra} observations. We  discuss features which could be detected with very deep observations with either \textit{Chandra} or future telescopes (Section~\ref{sec:Chandra_crazy}). We discuss model uncertainties, alternative KHI suppression mechanisms, and briefly compare to other CF clusters in Sect.~\ref{sec:discussion}. Section~\ref{sec:summary} summarizes our results.

\section{Method} \label{sec:method}
%
We perform idealized minor merger simulations where a gas free subcluster of $2\times 10^{13}M\subsun$ moved through the Virgo cluster from west to east, passing 100 kpc south of the cluster core  1.7 Gyr ago. \citet{Roediger2011} inferred this merger configuration as the most likely one from a detailed comparison of observed and simulated CF properties including orientation, distance to the cluster center, and contrasts in brightness and temperature across the fronts. Our simulations use the FLASH code (version 3.3, \citealt{Dubey2009}) and utilize the same method and initial conditions as described in \citet{Roediger2011} unless stated otherwise.

\subsection{Resolution}
We resolve the cluster core with a grid cell size of 0.5 kpc out to 110 kpc from the cluster center. 
The FLASH code captures the growth of the KHI even at a resolution of 16 grid cells per perturbation scale length (Roediger et al., in preparation, also Fig.~\ref{fig:extra} in the Appendix), hence we can resolve modes down to 8 kpc. We performed the same simulations at a coarser resolution of 1 kpc and found consistent results, but present the higher resolution results here. Further snapshots of a resolution test in the inviscid case can be found in Fig.~A1 of \citet{Roediger2011}. Convergence is to be expected for simulations at high viscosity, because viscosity damps the instability at small wavelengths first. Thus, as long as the  smallest unstable wavelength is resolved, increasing the resolution does not bring any new insights. At zero viscosity the simulations cannot formally converge, because the KHI grows fastest at small wavelengths. However, in the presence of multi-wavelength perturbations the largest modes which had sufficient time to grow dominate the visual appearance of a KH unstable interface.  The presence of smaller modes may add to the widening of the interface, which is captured in our simulations. Hence, our simulations sufficiently resolve the observable features. Naturally, the exact patterns of individual KH rolls will depend on the details of the seeding, which is inaccessible in any case.

\subsection{Viscosity}
We include a temperature-dependent dynamic viscosity $\mu=5500\,\mathrm{g}\,\mathrm{cm}^{-1}\mathrm{s}^{-1}f_{\mu}(T\ICM/10^8\K)^{5/2}$ (\citealt{Sarazin1988}, assuming a Coulomb logarithm of 40), where we parametrize the unknown level of viscosity suppression by the suppression factor $f_{\mu}$. 
Here our simulations differ from the ones presented by \citet{ZuHone2010}, who used a constant viscosity throughout the cluster. This may overestimate how effectively viscosity suppresses the
KHI because  it neglects the lower viscosity on the  cooler side of the CF that would apply for a
Spitzer-type viscosity. As pointed out above, the nature of the effective ICM viscosity is still unknown, and both models are valuable. In fact, the choice of \citet{ZuHone2010} was optimal for the focus of their work, which was to demonstrate the inability of viscous gas sloshing to prevent cooling flows, even in the presence of the highest reasonable viscosity. Using a temperature dependent viscosity provides a more stringent test of its ability to suppress the KHI, since having a lower viscosity on the cool side of the CF reduces viscous suppression of the KHI. Thus, this model is the better choice for our current purpose. Moreover, the Virgo cluster is much cooler than the cluster studied by \citet{ZuHone2010}, which might further reduce its effectiveness at suppressing KHI.

A full Spitzer thermal conductivity would erase temperature discontinuities rapidly: at a temperature of 3 keV and density of $3\times 10^{-3}\ccm$, the conduction timescale for temperature gradients with scale lengths of 3 kpc, i.e.~across the fronts, is about 0.3 Myr (\citealt{Sarazin1988}), much shorter than the age of CFs. 
Even if ordered magnetic fields prohibit the  conduction directly across the fronts, \citet{ZuHone2012} have shown that full Spitzer conductivity still erases the temperature discontinuity at the CFs in a few hundred Myr. The presence of temperature discontinuities at the observed CFs implies  strongly suppressed thermal conduction, hence we neglect it.

\subsection{KHI seeding}
The KHI needs seeds, initial perturbations, to grow. In contrast to earlier work we do not rely on seeding  by the discretization of the computational grid, but introduce intentional seeds by repeatedly adding small perturbations in the north-south velocity every 100 Myr, from pericenter passage on. The perturbation velocity is a superposition of sinusoidal waves of wave lengths 40, 20, 10, \ldots, 1.25 kpc with a total amplitude of $10\Kms$ at the first perturbation event. 
The amplitude is constant with wavelength.
The amplitude decreases linearly at subsequent perturbation events such that the perturbation is zero at 2 Gyr after pericenter passage. In this manner, we introduce a perturbation of a few percent or less of the typical shear velocity at each perturbation event, but keep the total  perturbation velocity well below the shear velocity. Perturbations of this level or higher can arise e.g.~from ICM turbulence due to AGN activity or the motion of galaxies (\citealt{Roediger2008,Vazza2012}). While our perturbations are highly idealized in their power spectrum and direction, they serve the purpose as seeds for the KHI. 

\subsection{Simulation runs}

We simulate the  merger history from 1 Gyr prior to pericenter passage to the fiducial time at 1.7 Gyr after the pericenter passage for the low viscosity case of $f_{\mu}=10^{-3}$. Here  KHIs occur from 0.5 Gyr after the pericenter passage on. We re-simulate the final 0.5 Gyr, i.e.~from 1.2 Gyr after the pericenter passage on, at higher viscosity with $f_{\mu}=0.01$ and 0.1. Besides reducing computational expenses, this procedure poses a more stringent test on the ability of the viscosity to suppress  KHIs because the instability is clearly developed at 1.2 Gyr. Additionally, we performed an inviscid control run.

\section{Suppression of KHI by viscosity} \label{sec:khsuppr}
%
\citet{Junk2010} derived a dispersion relation for the viscous KHI by standard linear perturbation analysis for the case of constant kinematic viscosity. It predicts reduced growth of the KHI for Reynolds numbers $<\pi\sqrt{\frac{(\rho\Cold+\rho\Hot)^2}{\rho\Cold\rho\Hot}}$, i.e.~for Reynolds numbers smaller than $\sim 10$ for moderate density contrasts of around 2 (\citealt{Roediger2012n7618}).  The shear velocity and the perturbation length scale enter the Reynolds number as the relevant velocity and length scale (Eqn.~\ref{eq:Re}). This agrees with the general physical expectation that viscous forces are equally important to inertial forces at a Reynolds number around 1. Thus for a ratio of 2 between the density on the cold side of the northern front, $\rho_{\rm cold}$, and that on the hot side of the front, $\rho_{\rm hot}$, in Virgo, a full Spitzer viscosity should affect perturbations of length scales smaller than $\sim 10\Kpc$. 

\begin{figure}
\rotatebox{90}{inviscid}
\rotatebox{90}{\phantom{"lghty"}}
\includegraphics[trim=280 0 600 0,clip,angle=90,width=0.32\textwidth]{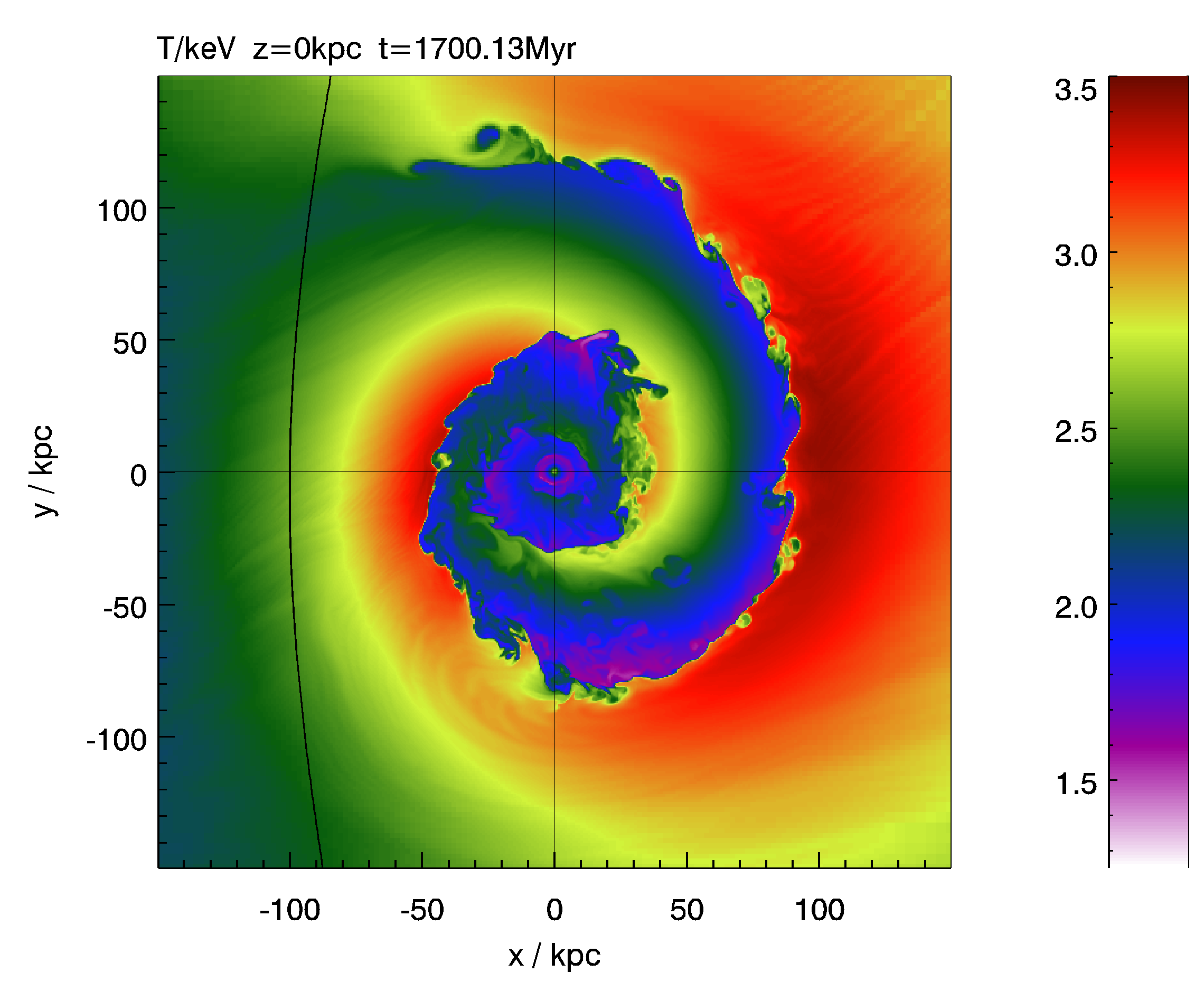}%
\includegraphics[trim=1860 250 0 0,clip,angle=0,height=4.2cm]{temp_sp00}
\newline
\rotatebox{90}{$10^{-3}$ Spitzer viscosity}
\rotatebox{90}{("low viscosity case" in text)}
\includegraphics[trim=280 0 600 0,clip,angle=90,width=0.32\textwidth]{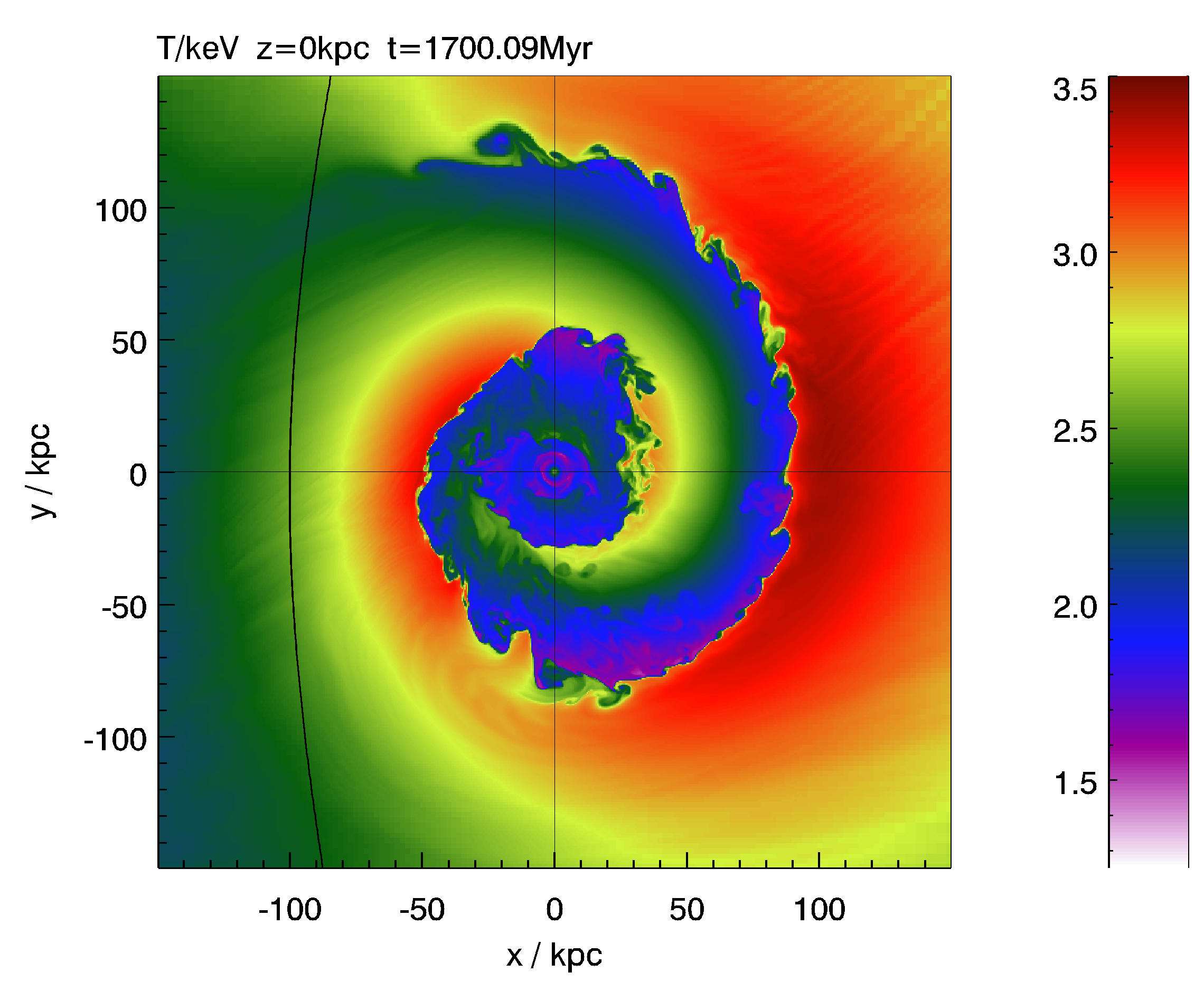}
\newline
\rotatebox{90}{$10^{-2}$ Spitzer viscosity}
\rotatebox{90}{\phantom{"lghty"}}
\includegraphics[trim=280 0 600 0,clip,angle=90,width=0.32\textwidth]{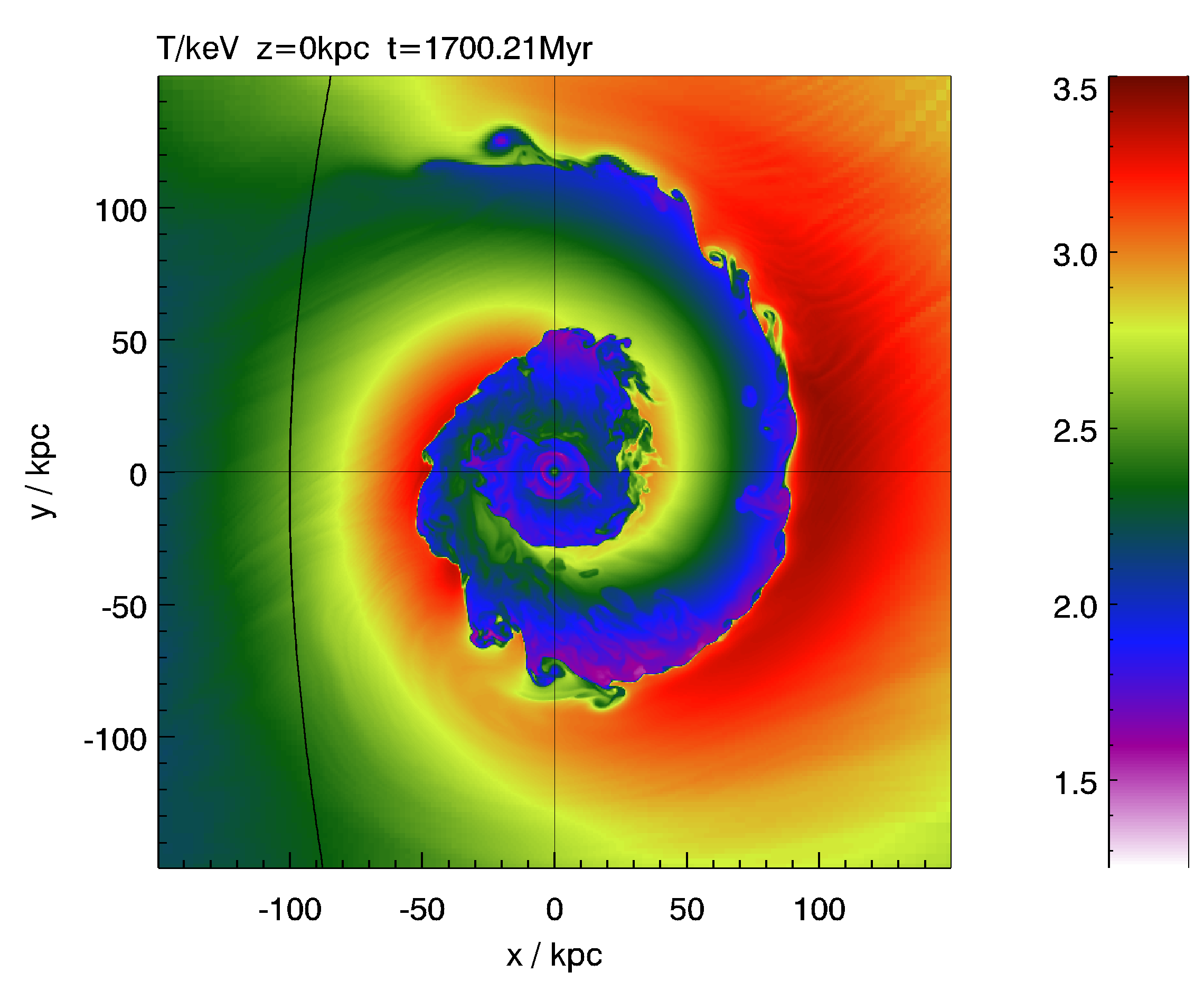}
\newline
\rotatebox{90}{\hspace{1cm}$0.1$ Spitzer viscosity}
\rotatebox{90}{\hspace{1cm}("high viscosity case" in text)}
\includegraphics[trim=50 0 600 0,clip,angle=90,width=0.32\textwidth]{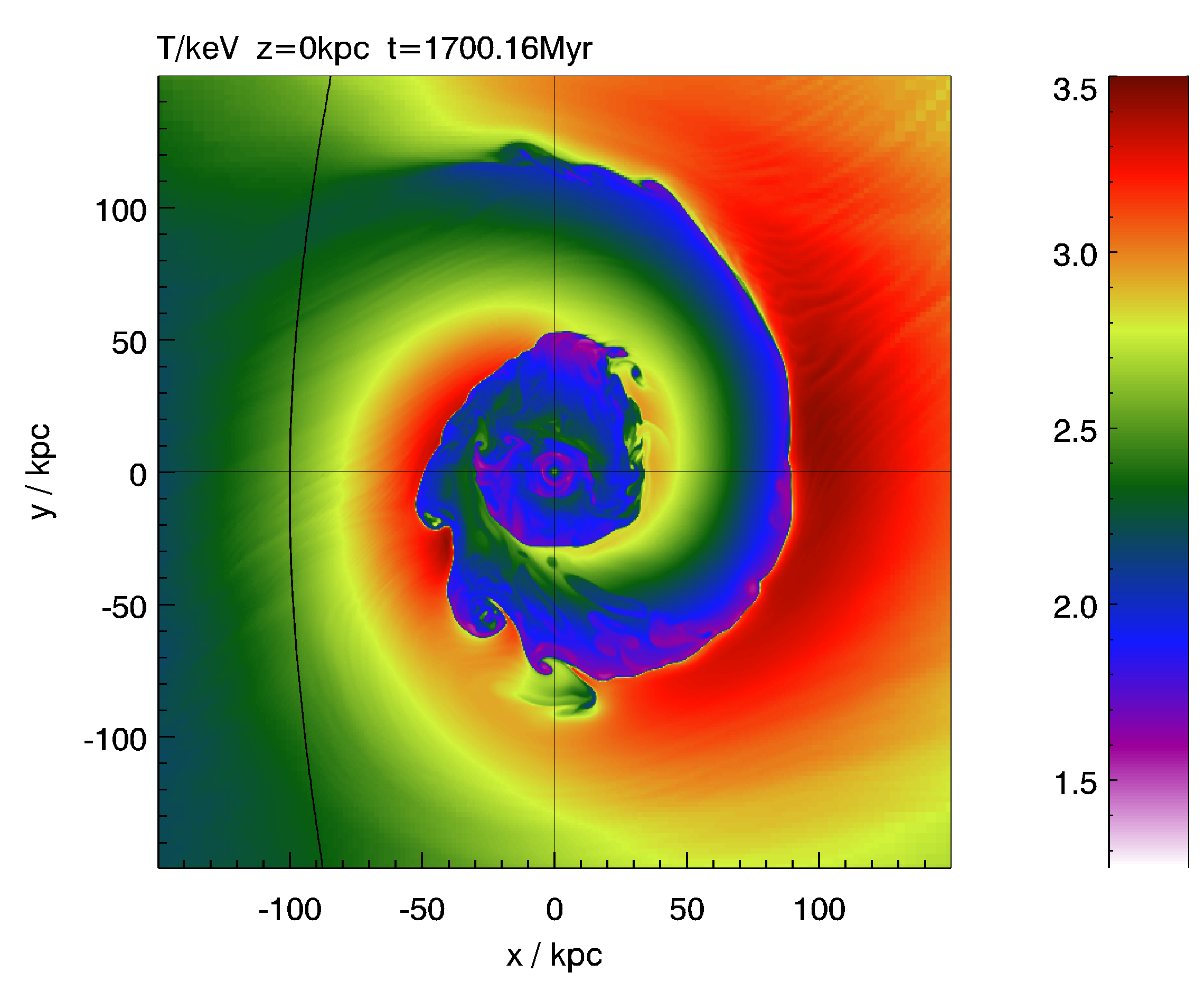}
\caption{Temperature slices in the orbital plane at the final timestep, for Spitzer-type, i.e.~temperature dependent, viscosities with suppression factors $f_{\mu}=0, 10^{-3}, 0.01$ and $0.1$ from top to bottom. Increasing the viscosity erases progressively larger substructure along the fronts. We have oriented the images such that they compare to the situation observed in Virgo, i.e.~north is up and west is right (see \citealt{Roediger2011} for details).}
\label{fig:tempslices}
\end{figure}

Figure~\ref{fig:tempslices} demonstrates the effect of the viscosity in the simulations. It displays temperature slices in the orbital plane at the final timestep for $f_{\mu}=0, 10^{-3}, 0.01$ and $0.1$ from top to bottom. 
At a viscosity $f_{\mu}\le10^{-3}$, all CFs are clearly distorted and made ragged by the KHI. With increasing viscosity, the fronts become less ragged, and structures at progressively larger scales are suppressed. Interestingly, even the small physical viscosity of $10^{-3}$ Spitzer erases some of the smallest perturbations present in the inviscid simulation. Finally, the fronts are almost completely smooth in the high viscosity case ($f_{\mu}= 0.1$) except for two large KH rolls separated by $\sim 40 \Kpc$ along the SW, where the shear flow is strongest ($\sim 500\Kms$). Distortions at smaller length scales though are absent at high viscosity at this location as well, whereas smaller distortions are present at this location at lower viscosity.  
%
%

Our simulations demonstrate that the viscosity is more efficient in suppressing the KHI than expected from the linear analysis. There are several reasons for this difference: the growth time is derived from the linear stability analysis. The effect of the viscosity is to reduce shear velocities, which applies to the flow parallel to the interface as well as the velocity perturbation in the perpendicular direction. Thus, while the linear analysis predicts only a slowed growth of the KHI, but still a growth, at longer timescales viscosity should shut off the growth completely and thus be more efficient than expected.  We demonstrate this behavior analytically and numerically in a separate publication (Roediger et al., in preparation). Furthermore, the sloshing CFs are curved interfaces embedded in a background gravitational potential and a stratified atmosphere, whereas the analytic estimate assumes a planar interface, no stratification and no gravity. The gravity at the northern CF suppresses KHIs at wavelengths above $\sim 50\Kpc$ and will thus slow down the growth of instabilities at somewhat smaller wavelengths, i.e.~the wavelengths seen in the simulations. Finally, the sloshing CFs are special contact discontinuities that are continuously reformed by the sloshing process, which interacts with the KHI and may modify its growth at long timescales. For example, the outwards motion of the CF stretches the relevant wavelengths, which reduces the growth rate and amplitude (\citealt{Churazov2004}).

\section{Observable features} \label{sec:observable}
%
\subsection{X-ray images}
\begin{figure}
\begin{center}
\rotatebox{90}{\phantom{lghty}}
\includegraphics[trim=0 0 0 0,clip,angle=0,width=0.45\textwidth]{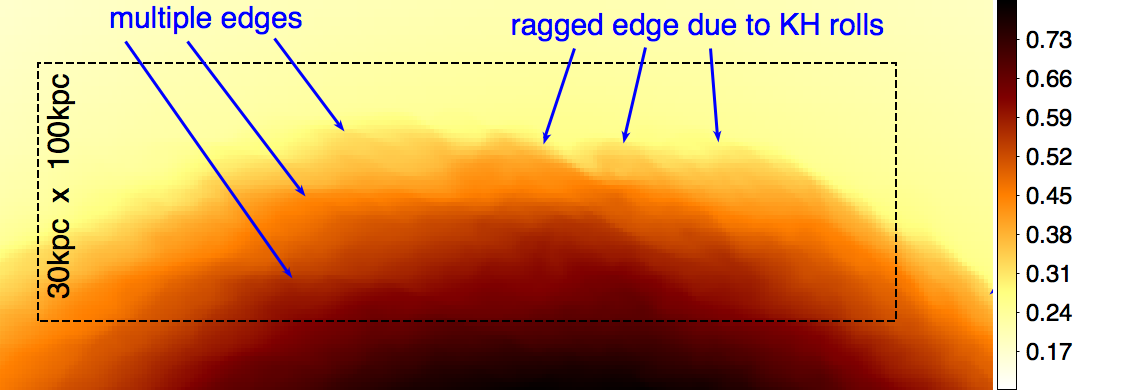}
\newline
\rotatebox{90}{$10^{-3}$ Spitzer viscosity}
\includegraphics[trim=0 100 0 0,clip,angle=0,width=0.45\textwidth]{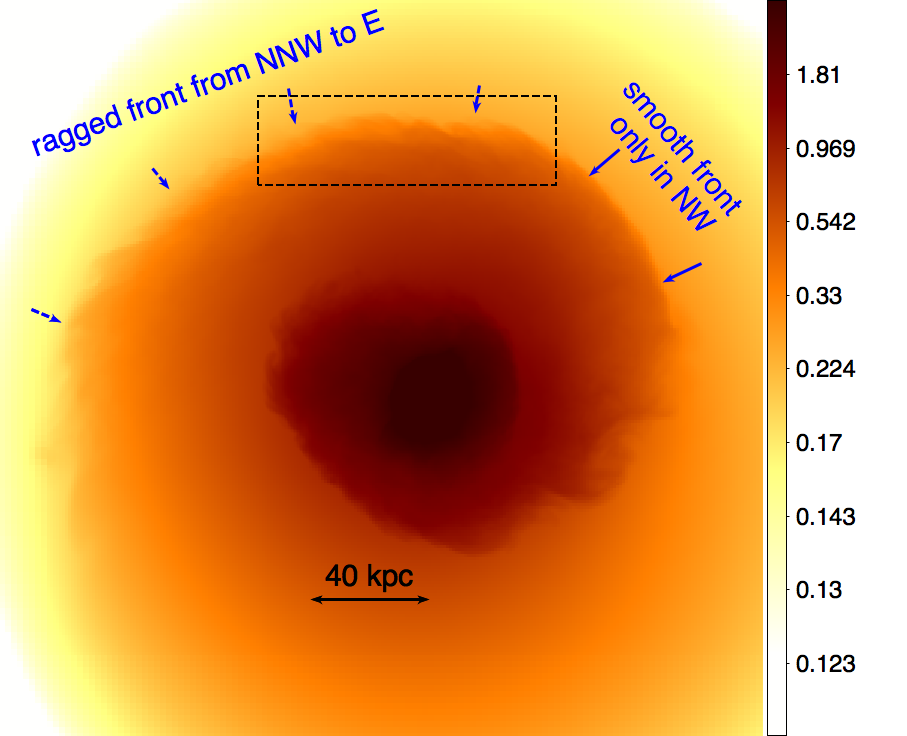}
\newline
\rotatebox{90}{0.1 Spitzer viscosity}
\includegraphics[trim=0 100 0 0,clip,angle=0,width=0.45\textwidth]{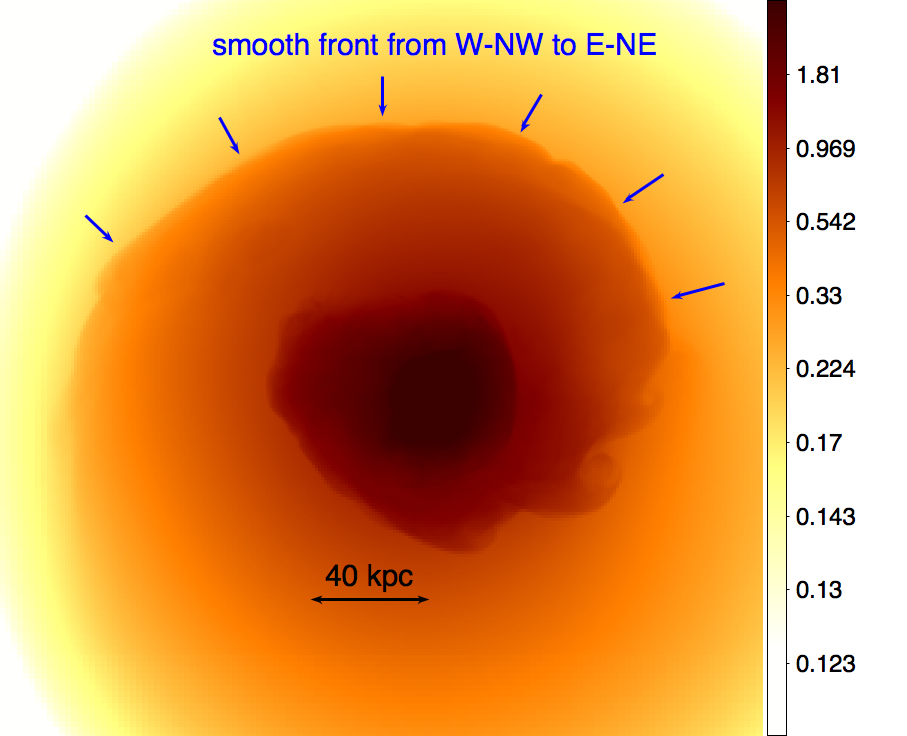}
\newline
\caption{Synthetic X-ray images in arbitrary logarithmic scale for the low ($f_{\mu}=10^{-3}$) and high viscosity ($f_{\mu}=0.1$) case in the middle and bottom panel. The top panel is a zoom-in in linear brightness scale on the northern part of the outer CF as marked by the dashed rectangle. Prominent structures are labelled.}
\label{fig:Xray}
\end{center}
\end{figure}
%
We calculate synthetic X-ray images by projecting $n\ICM^2\Lambda(T\ICM)$ along the line-of-sight (LOS), where $\Lambda(T)$ is the cooling function according to \citet{Sutherland1993} and we assume a metallicity of 0.3 solar. Figure~\ref{fig:Xray} displays the predicted X-ray images for the viscosity suppression factors $f_{\mu}= 10^{-3}$ and $f_{\mu}= 0.1$. We will refer to these two viscosities as low and high viscosity, respectively.

As the inner CFs are likely to be distorted by the AGN activity in the Virgo cluster center (\citealt{Forman2007}), we focus on the structure of the outer northern front.
The shear flow along this front is weakest in the NW, leading to a smooth, sharp front in the NW independent of viscosity. Along the north (N) and the east    of the front the shear flow is stronger ($\sim 300\Kms$)   and forms distinct structures depending on the viscosity (Fig.~\ref{fig:Xray}). At high viscosity, the front forms a smooth arc here as well, but it is ragged at low viscosity.  Individual KH rolls at $\sim 15 \Kpc$ size can be identified as triangular-shaped irregularities. They give the front a saw tooth like appearance.  A further characteristic pattern is multiple adjacent brightness edges with a spacing of about $5\Kpc$ parallel to the main front.  We have labelled these features in the  zoom-in in the top panel in Fig.~\ref{fig:Xray}.

\subsection{Brightness profiles}
Even though the CFs are discontinuities in density and temperature, they do not lead to true discontinuities in surface brightness, but in its slope. The X-ray brightness profile of a spherical cloud with a  power-law density profile can be approximated by $S_X(d) \propto \sqrt{-d}$ near its edge (\citealt{Vikhlinin2001}). In our notation, the coordinate $d$ increases with increasing distance from the cluster center and is defined to be zero at the CF, i.e.,  $-d$ is the distance to the projected edge measured inwards. The minus-sign is required because the brightness decreases with increasing distance to the center. 
%
\begin{figure}
\centering\includegraphics[trim=0 100 0 20,clip,angle=0,width=0.3\textwidth]{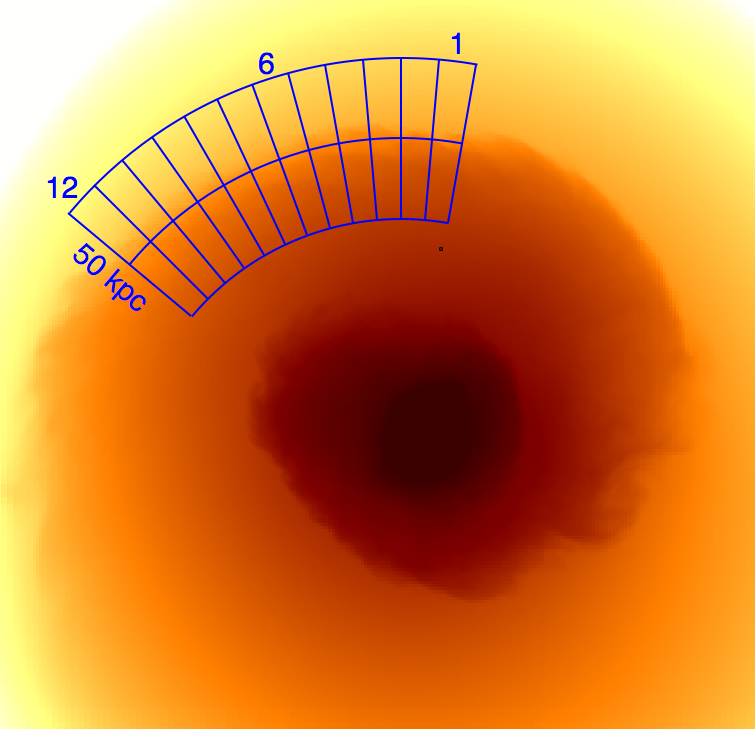}\newline
\centering\includegraphics[angle=0,width=0.4\textwidth]{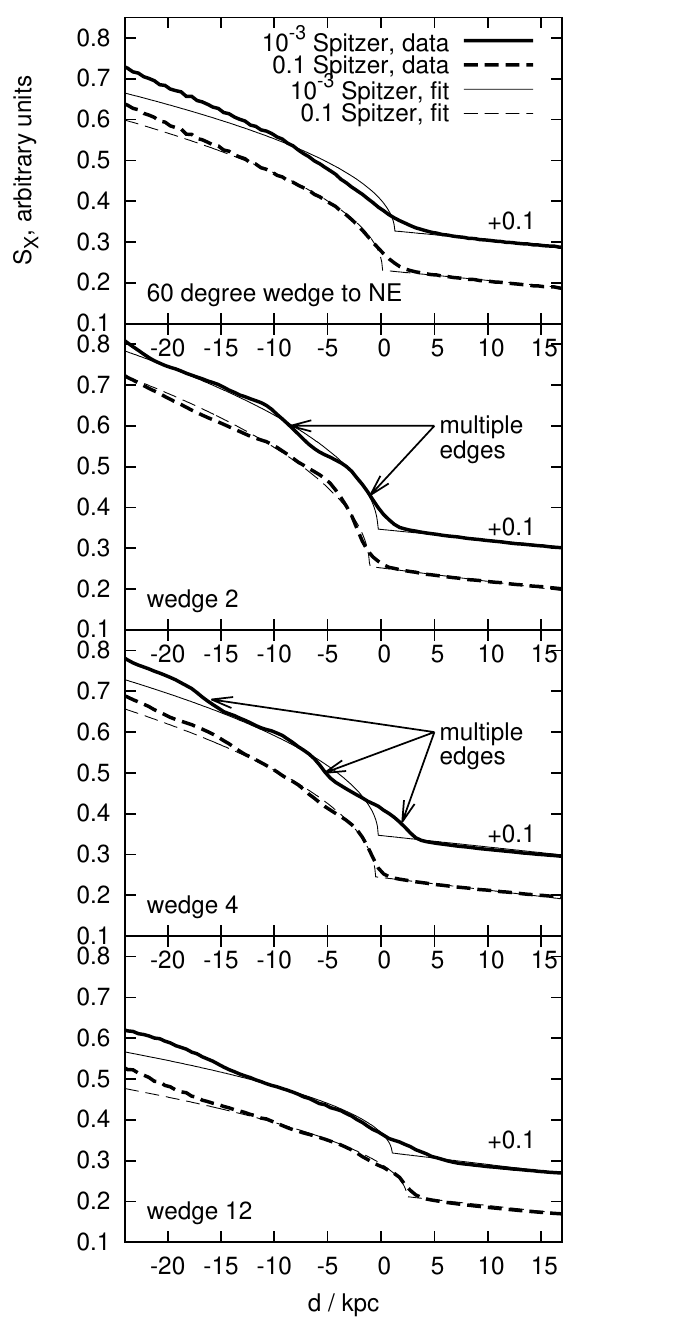}
\newline
\caption{X-ray surface brightness profiles across the outer CF towards the NE. The extraction wedges are shown in the top panel. The individual wedges 1 to 12 are $5\degree$ wide. The first line plot compares profiles for the total $60\degree$ wedge in the high and low viscosity case, see legend. For clarity, we have offset both cases by 0.1. We fit each profile by the expected brightness profile for a broken power law (Eqn.~\ref{eq:sqrt_d}).   Thick lines are for the simulation data, thin lines are the corresponding fits. The bottom three panels show the same comparison for selected narrow wedges as labelled in each panel.}
\label{fig:profiles}
\end{figure}
%
Figure~\ref{fig:profiles} compares brightness profiles across the CF towards the north-east (NE) for  low and high viscosity. We extract  profiles in narrow, $5\degree$ wedges, corresponding to a width of 9 kpc along the front, and for a wide wedge of $60\degree$ (see top panel in Fig.~\ref{fig:profiles}). The vertex of the wedges is not the cluster center, but we aligned the annuli of the wide wedge with the CF. While this makes the detailed interpretation of the profiles more difficult in terms of assumed geometry, this choice optimizes the sharpness of the edge in the profiles, which is the main purpose here. 
In all cases the CF appears as a steeper slope in the brightness profiles.  
We fit the profiles within 15 kpc around the edge with the expected broken power-law profile. More precisely, we allow the position of the edge $d\CF$ to be a free parameter by working in the offset coordinate $x=d+d\CF$ along the profile, and include a linear background profile outside the CF:
\begin{eqnarray}
S_X(d) &=& S(x - d\CF) \nonumber \\
&=& 
\left\{
	\begin{array}{ll}
		A_0\sqrt{(-x+d\CF)/\Kpc}+S_0 \; & \mbox{if } x < d\CF, \\
		 m(-x+d\CF)+S_0 & \mbox{if } \tilde x \ge d\CF,
	\end{array}
\right.
\label{eq:sqrt_d}
\end{eqnarray}
The amplitude of the jump $A_0$, the position of the edge $d\CF$,  the background brightness $S_0$, and the slope of the background profile $m$ are free parameters.  We have performed the same exercise for profiles towards the NW  (not shown here), where the CF appears undistorted in the surface brightness images, and find good agreement with the expected $\sqrt{-d}$ shape. In the high viscosity case,  profiles across the NE edge agree well with this expectation as well (Fig.~\ref{fig:profiles}). In the low-viscosity case, however, profiles across the ragged NE  edge show a distinct deviation from the $\sqrt{-d}$ profile. The slope at the edge is shallower, the edge appears washed or blurred out. In  many of the narrow wedges, the profiles display multiple edges, corresponding to the multiple adjacent fronts seen in the image.

\section{Current and future observations} \label{sec:virgo_newobs}

\subsection{Evidence for KHIs in existing data} \label{sec:evidence}

The northern CF in the Virgo cluster, as evident in the  \textit{XMM-Newton} mosaic around M87, is highlighted by arrows in  Fig.~\ref{fig:virgocf}. While this CF appears sharpest, i.e.~has the steepest brightness gradient, in the N-NW, it continues in a somewhat diffuse manner to the NE (see also brightness residual map in Fig.~1 in \citealt{Roediger2011}). This matches the predictions of our simulations at low viscosity. Due to the absence of a shear flow along the NW, the CF should be sharp here. Along the N and NE, our simulations show a stronger shear flow and consequently KHIs at low viscosity. In this shallow \textit{XMM-Newton} observation, individual KH rolls cannot be detected, nor their absence confirmed, because of several factors. These include the broad-winged PSF of \textit{XMM-Newton}, the limited counting statistics of the data (6.5 ks of good time for the PN camera after flare filtering in this region of the CF), and the complex non-uniform detector response and background.

\begin{figure}
\centering\includegraphics[trim=0 00 0 00,clip,width=0.45\textwidth]{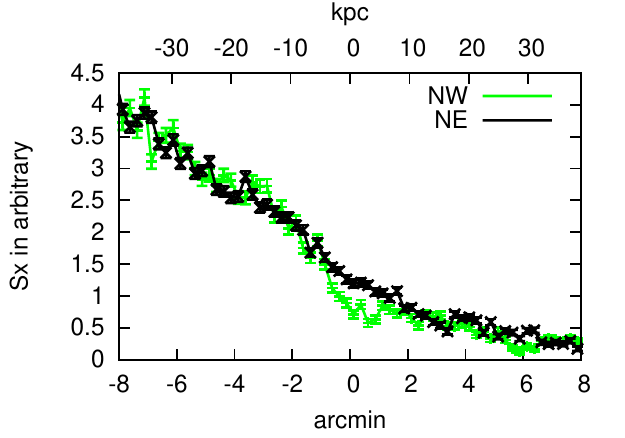}
\caption{Surface brightness profiles across the NW (green/light grey in print) and NE (black) regions of the Virgo cluster sloshing CF, extracted from the archival \textit{XMM-Newton} data.  The zero point on the x axis denotes the approximate position of the surface brightness edge of the CF.  The scaling of the surface brightness is arbitrary. The profile wedges cover about 5 arcmin or 24 kpc along the CF.  Note the sharp rise of the surface brightness behind the CF in the NW and the slower rise at in the NE. 
}
\label{fig:cf}
\end{figure}

To quantitatively confirm the visual impression of the smeared-out front in the NE we extracted surface brightness profiles from the archival \textit{XMM-Newton} data across the CF both in the NW and the NE. Background was estimated using the dark sky background files, and the ODF files were reprocessed to incorporate the most up to date calibration data and filtered for periods of high background.  We chose profile wedges which cover about 5 arcmin (24 kpc) along the CF and carefully aligned the annuli of these wedges with the CF such that the radius of curvature of the sector matched that of the CF.  The profiles are shown in Fig.~\ref{fig:cf}.   In the NW, the brightness profile exhibits a steep rise as expected for an underlying density discontinuity. In the NE, however, the brightness rises less sharply, indicating a smeared out front as expected for a  low ICM viscosity. A similar apparent smearing  may arise from a mis-alignment of the profile annuli with the front. We experimented with different extraction regions and could not get a sharper edge in the NE than shown in Fig.~\ref{fig:cf}. Hence, the blurred CF in the NE suggests the presence of KHIs and thus strongly suppressed viscosity in the ICM. Individual KH rolls may be visible in a deep \textit{Chandra} observation.

Our simulated X-ray images exhibit two prominent KH rolls in the SW of the cluster core at both high and low viscosity. If present in this evolutionary stage, they might be detectable in the current \textit{XMM-Newton} image.  However, the exact shape and height of  the KH rolls depends on the details of their seeding, which includes a full turbulent spectrum and perturbations from the central AGN outbursts in the real Virgo cluster, which is more complex than in our simulations. Furthermore, the rolls are less clear at a somewhat less optimal orientation of the LOS with respect to the merger plane. Thus, the basic prediction of the simulations is that the CF in the SW is especially susceptible to the KHI, and the \textit{XMM-Newton} image indeed shows a less sharp front in this region. We  note that these large KH rolls appear less clear in the X-ray images at \emph{low} viscosity, because their borders are washed out by smaller-scale instabilities. At a viscosity of 0.1 Spitzer, these rolls appear much clearer and might be visible in the shallow \textit{XMM-Newton} data, but they are not. Instead, the CF in the SW simply appears blurred out over $\sim 20\Kpc$, as would be expected at low Virgo ICM viscosity. Thus, the structure of both the NE and the SW fronts suggests a strongly suppressed ICM viscosity.

\subsection{Detectability of Kelvin-Helmholtz Instabilities in deep \textit{Chandra} observations} \label{sec:detect}

\subsubsection{Direct imaging}

\begin{figure*}
\includegraphics[width=\textwidth]{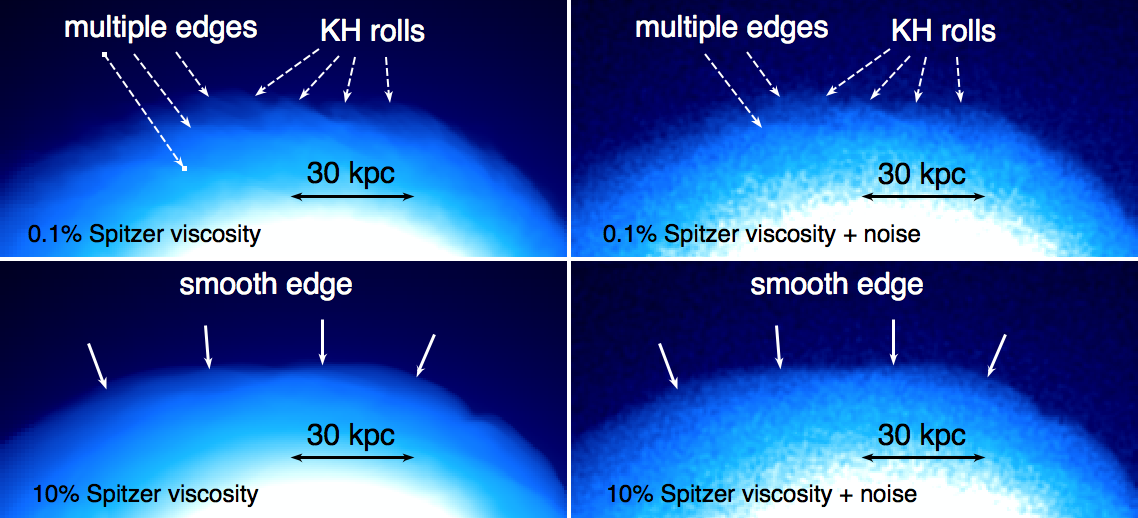}
\caption{Simulated X-ray images  of the northern sloshing CF in the Virgo cluster at different viscosities. The top and bottom rows are for low and high viscosity ($10^{-3}$ and $0.1$ of the Spitzer value), respectively. The left-hand-side column shows noiseless images, in the right-hand-side column we added a random Poisson deviate to match the surface brightness and noise level of a simulated 300 ks \textit{Chandra}/ACIS-I observation. The structure of the CF  differs between low and high viscosity.  The KHIs can be clearly seen in the former case (see labels), in both the ideal and in the noisy image.}
\label{fig:sim}
\end{figure*}

The left panels in Fig.~\ref{fig:sim} show the direct comparison of  surface brightness images at high and low viscosity  along the northern sloshing CF of the simulated Virgo cluster, i.e.~a smooth front in the high viscosity case and a ragged front at low viscosity. This field of view corresponds to two ACIS-I pointings of the \textit{Chandra} X-ray observatory. 

To evaluate the detectability of these structures in real observations, we match the count density in the simulated images to the surface brightness measured for the Virgo cluster CF in the  \textit{XMM-Newton} exposure. Using PIMMS, we scale it to a 300 ks \textit{Chandra}/ACIS-I observation. With this exposure time, there will be $\sim$100 source counts per 0.5 kpc$\times$0.5 kpc pixel (6$''\times$6$''$) just behind the CF.  A random Poisson deviate is then added to each 0.5 kpc$\times$0.5 kpc pixel to simulate the noise of a real observation.   The resulting noisy images are shown in the right column of Fig.~\ref{fig:sim}. We neglect background in these idealized simulations, because the count rate from the background is $<$10\% of the rate from the gas inside the cold front and thus insignificant.  

The KH rolls in the low viscosity case are clearly visible in the data with the random deviate added, and they are again absent in the high viscosity case.   The KH rolls span spatial scales of several kpc.  The variations in surface brightness due to the KH rolls in the low viscosity simulation are as large as $\sim$20\%.  There are $\sim$4000 counts in a 5 kpc $\times$ 2 kpc region behind the cold front in the simulated \textit{Chandra} image. Thus we could in principle detect variations in surface brightness of $\sim$5\% at 3$\sigma$ confidence with a real observation.  The KH rolls on scales of a few kpc, if present, would be easily detectable at more than 10$\sigma$ confidence in such a deep \textit{Chandra} observation.  Additionally, we could  observe KH rolls on smaller scales at lower significance.  Twenty percent variations of surface brightness could be detected at 4$\sigma$ confidence in 1 kpc$\times$1 kpc scale regions.

\subsubsection{Profiles}

\begin{figure}
\centering\includegraphics[trim=0 0 0 0,clip,width=0.35\textwidth]{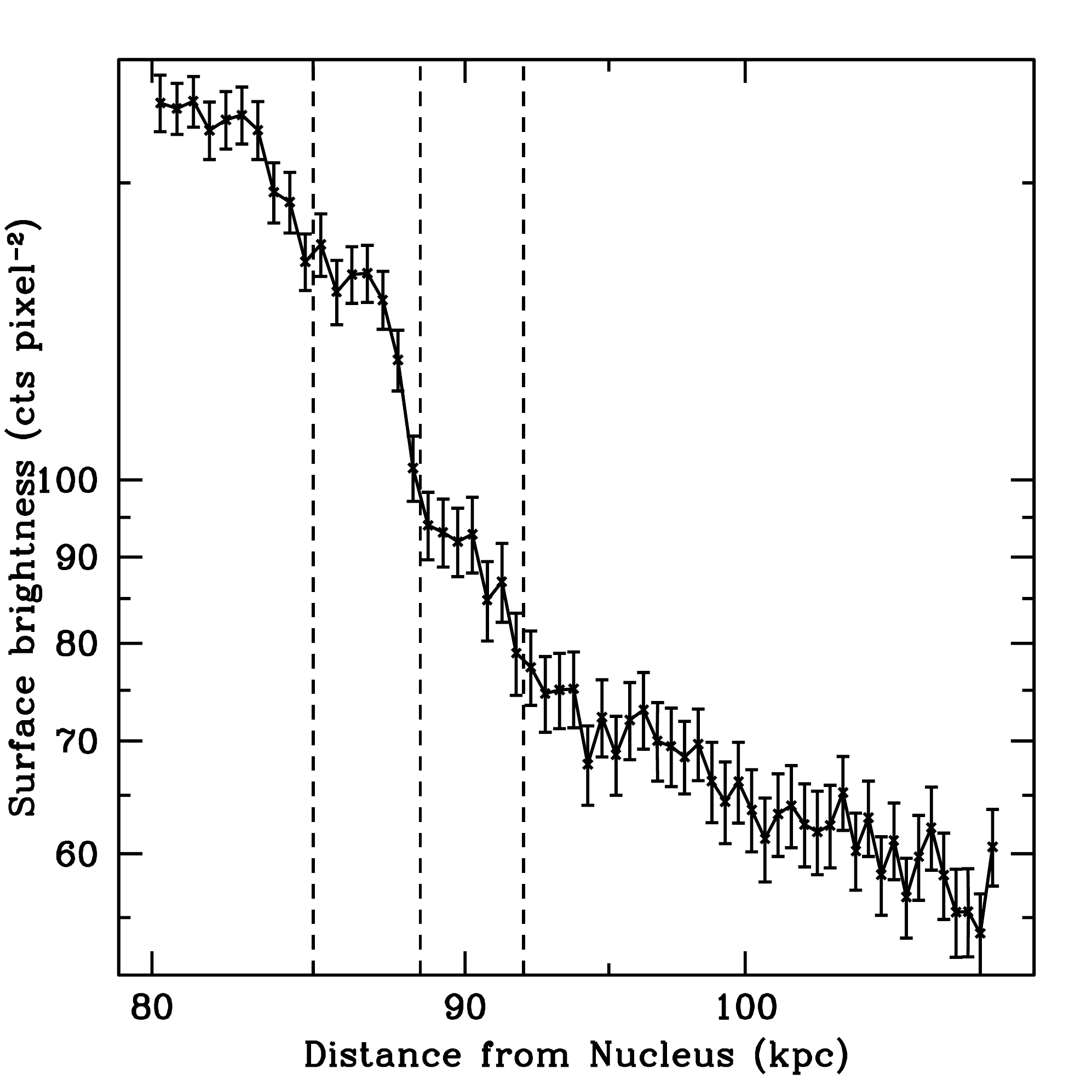}
\newline
\centering\includegraphics[trim=0 0 0 0,clip,width=0.35\textwidth]{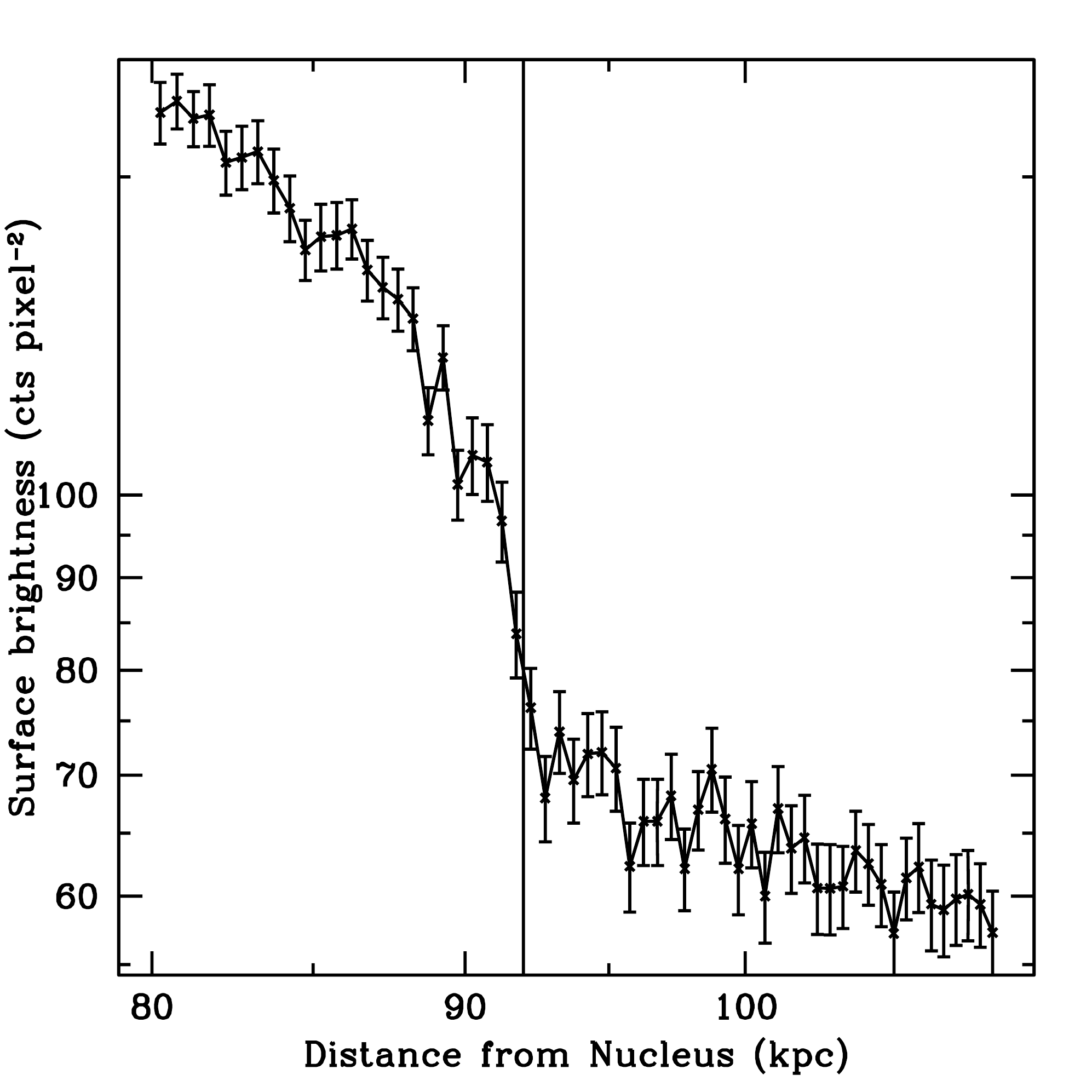}
\newline
\caption{Surface brightness profiles in 1.5$^\circ$ wedges (0.5 arcmin or 2.5 kpc at the CF) across the cold front in the  low (top - 10$^{-3}$ Spitzer) and high (bottom - 0.1 Spitzer) viscosity cases.  The  CF is at $\sim$90 kpc.  Note that the surface brightness profile of the low viscosity simulation shows multiple edges denoted by the vertical dashed lines.  Such discontinuities are indicative of density jumps. They arise due to the KH rolls.  
The surface brightness profile for the high viscosity case displays only a single steep edge, indicating the absence of KH rolls.}
\label{fig:sbp}
\end{figure}

To further demonstrate the detectability of the KH rolls in the simulated data, we extracted surface brightness profiles across the CF in 1.5$^\circ$ wedges in both data sets with the random Poisson deviate added.  Two examples are shown in Fig.~\ref{fig:sbp}. We follow the classic observational data analysis strategy and fix the vertices of the wedges at the cluster center.  The wedge opening angle of 1.5$^\circ$ corresponds to a linear distance of 2.5 kpc or 0.5 arcmin at the CF and is thus much narrower than in our analysis of the \textit{XMM-Newton} data.  Despite the added Poisson noise the predicted multiple edges in the low viscosity case can be clearly detected and are marked by vertical lines.  The spacing between  the surface brightness edges (i.e.~the spacing between the vertical lines - 3 to 4 kpc) is roughly the height of the KH rolls.  The height of these rolls is typically a third or half of the scale length of the KH rolls.  Thus the edges spaced roughly 4 kpc apart in the low viscosity profile signify KHI scale lengths of $\sim$10 kpc.  The presence of these edges in a real observation would immediately give the typical scale length of KH rolls, and these rolls should be present if $\nu_{\rm ICM} \ll \nu_{\rm Spitzer}$. In contrast, the surface brightness profile of the high viscosity simulation can be well described with a single power law with no evidence of an edge or change in slope inside the contact discontinuity of the CF.

\subsubsection{Statistical characterization}
At low viscosity, there is a wealth of structure inside the CF.   We created the power spectral density (PSD) function of the emission behind the CF in  both cases to search for a signature in the Fourier domain of the spatial scales where the KH rolls are present.  The PSD is indeed larger in the low viscosity simulation on scales where the KH rolls are present, but  we could not directly recover the spatial scales at which the KHIs were induced.  We can confirm the presence of the KH rolls using the PSD, but we do not expect to be able to extract the specific spatial scales of the KH rolls in Fourier space. 
As an alternative, we produced brightness profiles \emph{along} the CF and derived their 1D power spectra. In a  narrow annulus along the CF the KH rolls appear as ripples in the azimuthal profile. To collect a sufficient number of counts, the annulus needs a certain width, but increasing the width of the annulus tends to average out the surface brightness wriggles due to the saw-tooth pattern of the KH rolls. As a consequence, the 1D power spectrum analysis does not reveal more information than can be obtained by visual inspection of the images. The Fourier space analysis is generally better suited for (quasi-)periodic signals, but we have only  a few KH rolls in the simulated ACIS-I image.

\subsection{Structures in very deep observations}  \label{sec:Chandra_crazy}

\begin{figure*}
\includegraphics[width=\textwidth]{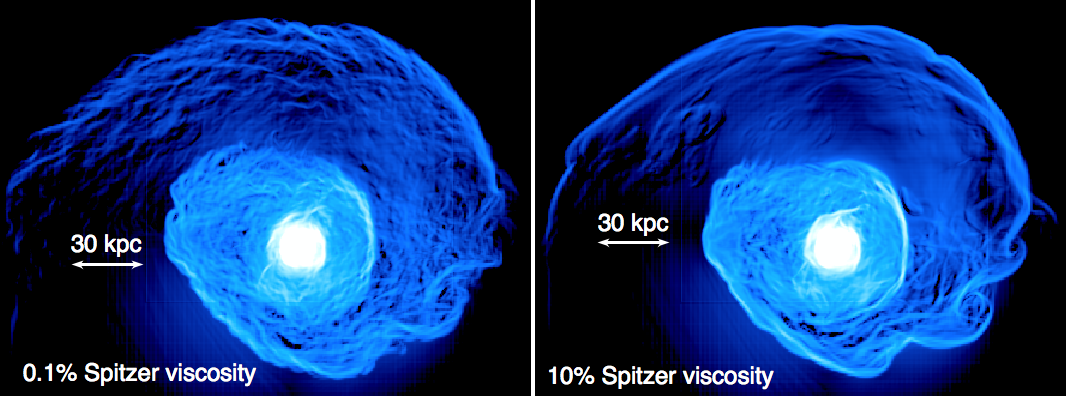}
\caption{
Simulated images of the Virgo cluster CF with Sobel filter applied, low viscosity on the left, high viscosity on the right. This filter selects regions of steep brightness gradients, i.e.~edges. Note the difference in the structure along the  CFs in the two images, and the clear evidence of turbulent flow behind the CF in the low viscosity (i.e., high Reynolds number) case.}
\label{fig:sobel}
\end{figure*}

The differences between high and low viscosity sloshing are more clearly demonstrated using an edge detection algorithm.  We applied a Sobel filter, a discrete differentiation operator often used for edge detection, to both simulated surface brightness images as shown in Fig.~\ref{fig:sobel}.  These filtered images were created using the raw (i.e., noiseless) simulated image and represent very deep observations by \textit{Chandra} or a future large area, low background X-ray observatory where the noise from counting statistics is small.  There are several features of note.  First, the sloshing cold front is a sharp, continuous arc in the high viscosity simulation, whereas the CF is distorted in the low viscosity case.  The approximate scales of the KHIs can be clearly  seen in the filtered, low viscosity image.  Second, the flow of the gas interior to the CF in the low viscosity case is turbulent, evident from the wealth of structure inside the cold front. Even in the presence of significant Poisson noise, these surface brightness fluctuations may be detectable in the power spectral density of fluctuations of the surface brightness in this region.

\section{Discussion} \label{sec:discussion}
%

\subsection{Model uncertainties}
In this section, the impacts of various assumptions on the conclusions are discussed.

\subsubsection{Merger history}
The cluster model and the assumed merger history determine the dynamical parameters relevant for the growth of the KHI at the CFs, namely the local density, temperature, their contrasts across the front, and the amplitude of the shear velocity.  The density,  temperature and their contrasts across the CF are constrained observationally and matched in our simulation, hence these quantities are sufficiently accurate. The amplitude of shear velocity will depend somewhat on the merger strength. However, our results regarding the growth or suppression of the KHI are not sensitive to the exact shear velocity within a factor of 1.5. The shear velocity is unlikely to be outside this interval, because this implies either a much stronger or much weaker merger. The former would lead to larger contrasts of brightness and temperature across the fronts than observed and an overall front distortion of the sloshing structure. The latter would lead to too weak contrasts or even the absence of a true discontinuity at the northern front (\citealt{Roediger2011}). Neither case is observed. Moreover,  a viscosity of only 0.1 of the Spitzer value suppressed the KHI along the northern CF in our fiducial merger. A full Spitzer viscosity will certainly suppress the KHI in a moderately stronger merger.
Hence our conclusions are little affected by uncertainties in the merger history.

\subsubsection{Line-of-sight (LOS)}
%
\begin{figure*}
\hspace{2cm} $f_{\mu}=10^{-3}$ \hfill $f_{\mu}=0.1$ \hfill\phantom{x}\newline
\rotatebox{90}{$i=0$, LOS $\perp$ to orbital plane}
\includegraphics[trim=300 0 380 100,clip,angle=90,width=0.41\textwidth]{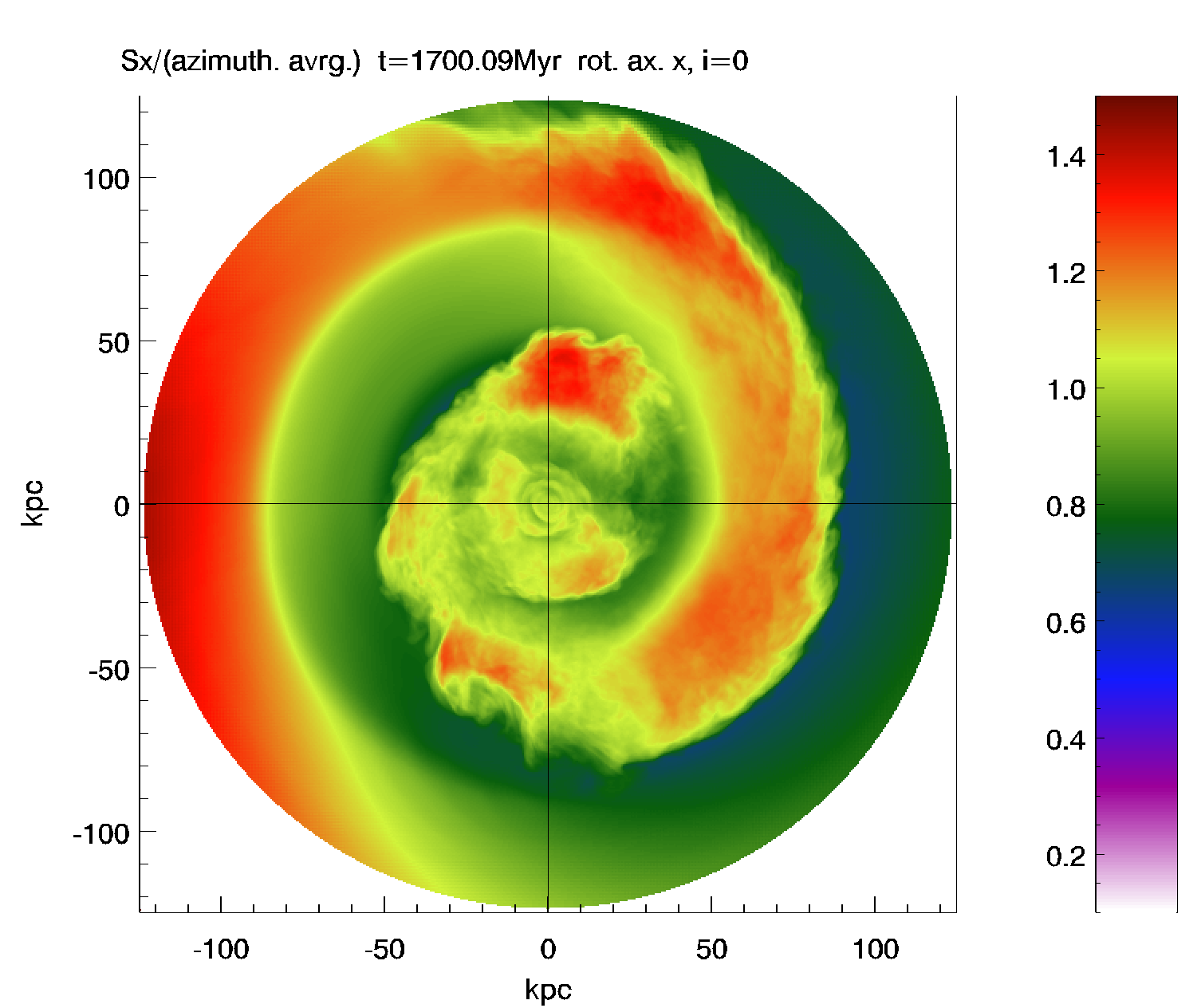}
\includegraphics[trim=300 0 380 100,clip,angle=90,width=0.41\textwidth]{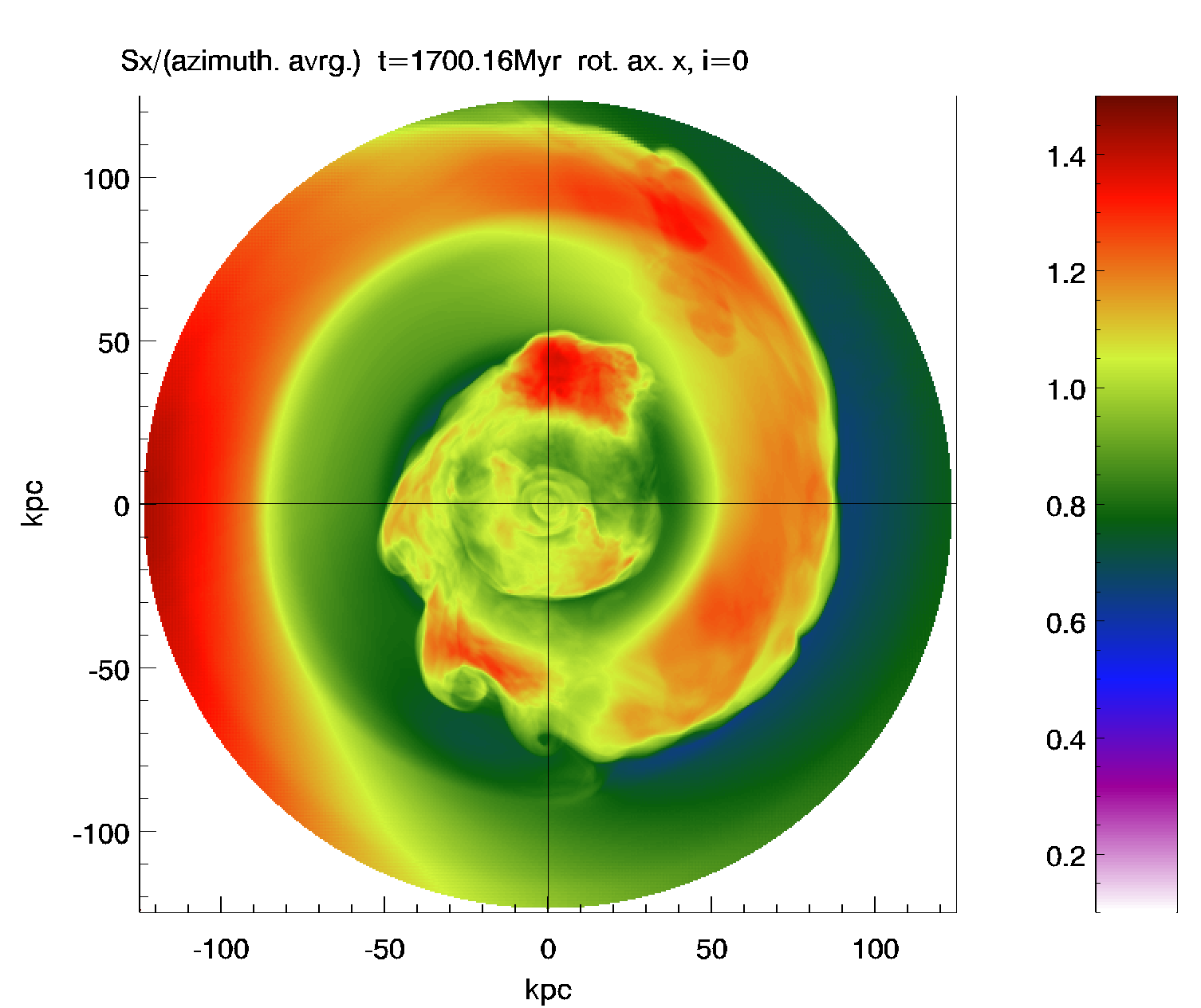}
\includegraphics[trim=1300 100 0 100,clip,angle=0,height=5.5cm]{viscangleResiduals_ax_x_ang_0}
\newline
\rotatebox{90}{\hspace{2cm}$i=30$}
\includegraphics[trim=300 0 380 100,clip,angle=90,width=0.41\textwidth]{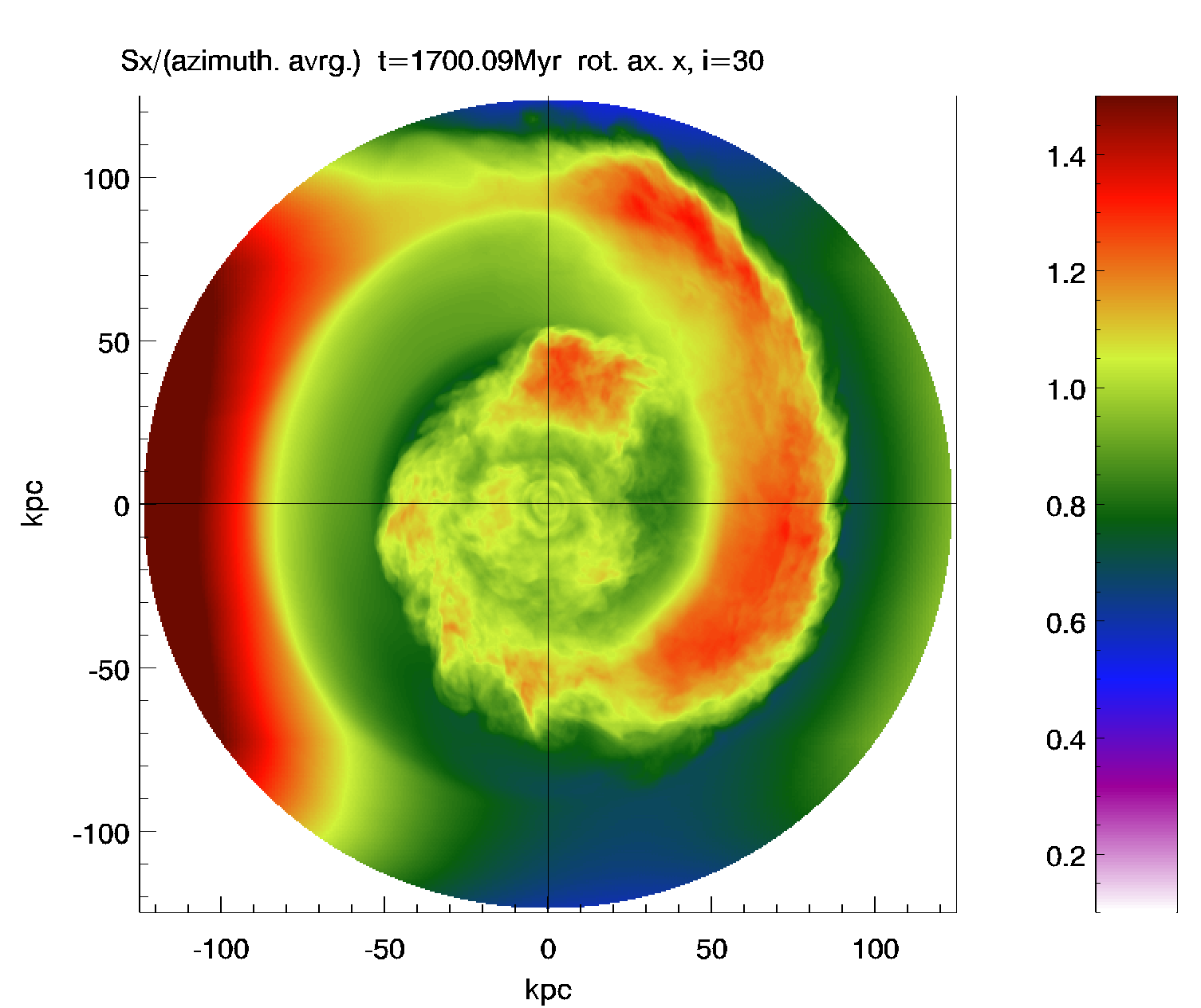}
\includegraphics[trim=300 0 380 100,clip,angle=90,width=0.41\textwidth]{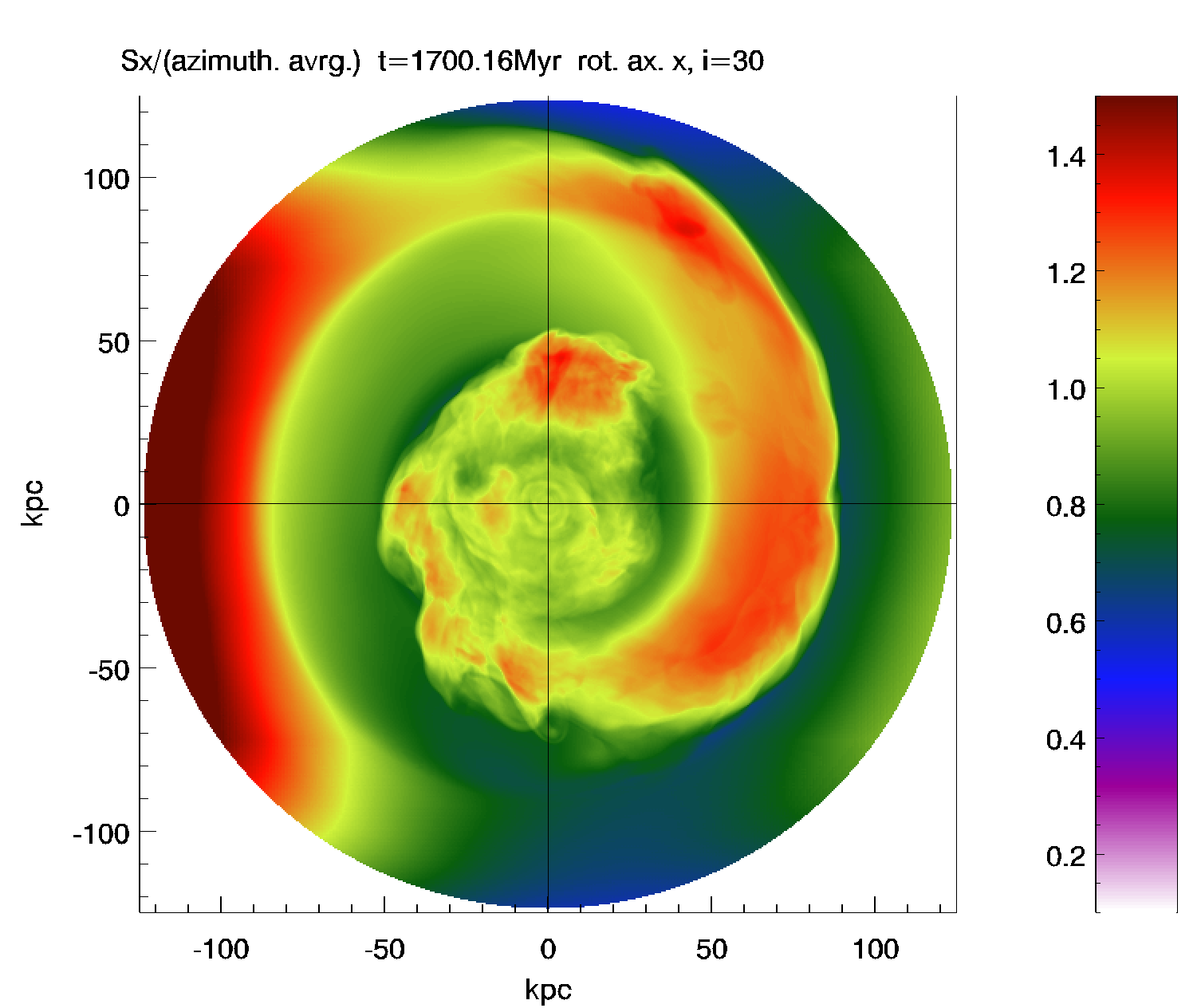}
\newline
\rotatebox{90}{\hspace{2cm}$i=60$}
\includegraphics[trim=300 0 380 100,clip,angle=90,width=0.41\textwidth]{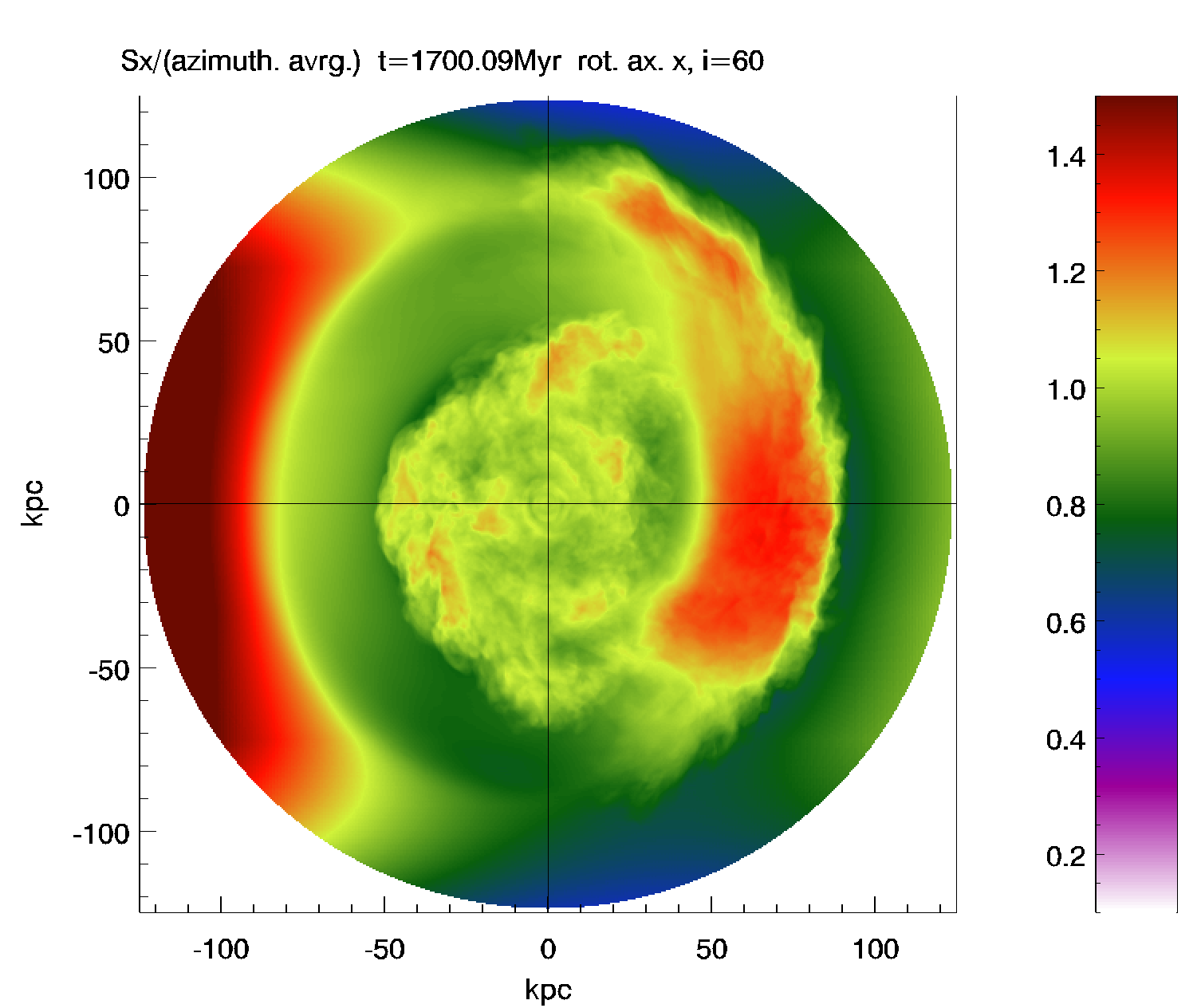}
\includegraphics[trim=300 0 380 100,clip,angle=90,width=0.41\textwidth]{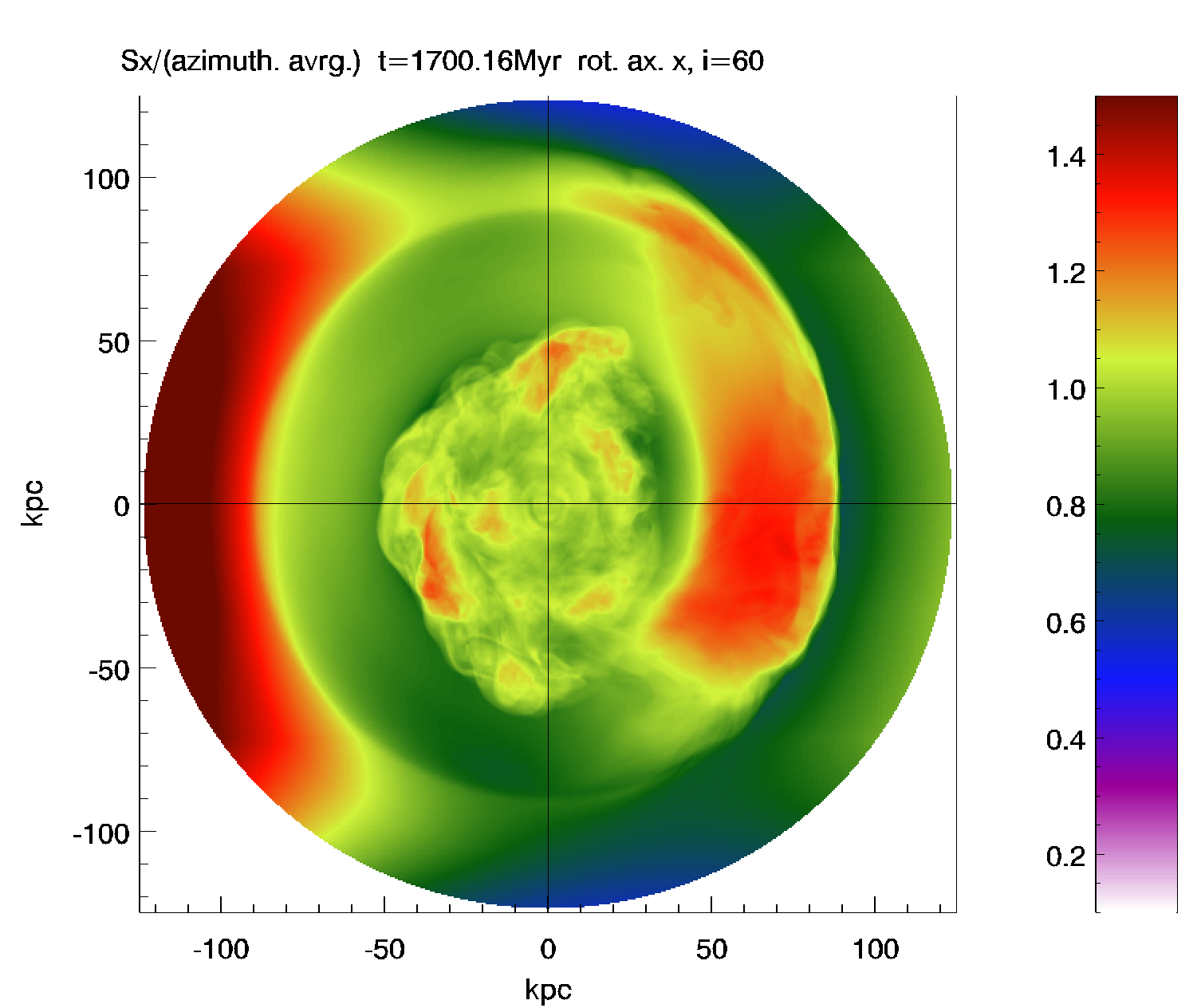}
\newline
\rotatebox{90}{$i=90$, LOS in orbital plane}
\includegraphics[trim=300 0 380 100,clip,angle=90,width=0.41\textwidth]{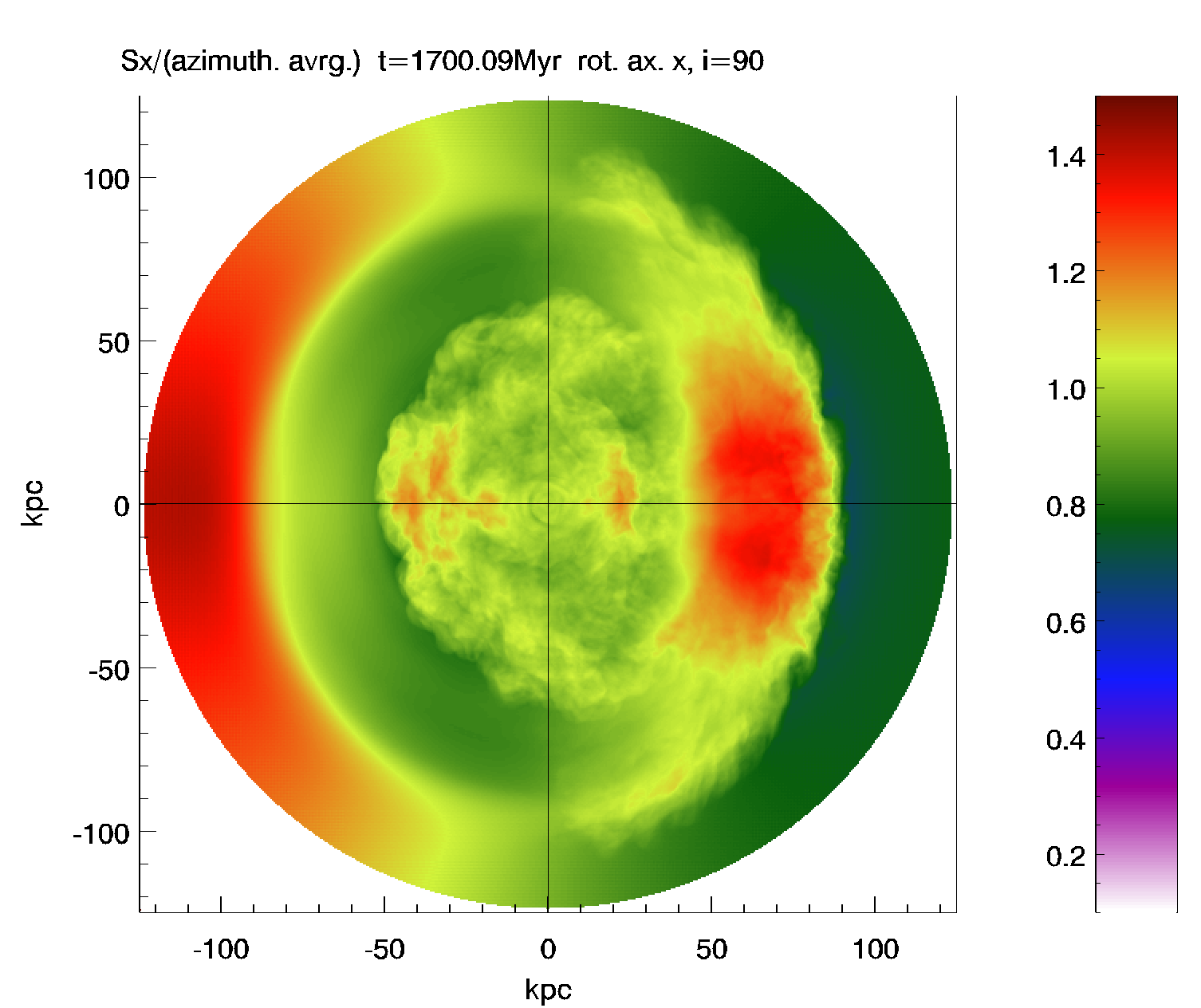}
\includegraphics[trim=300 0 380 100,clip,angle=90,width=0.41\textwidth]{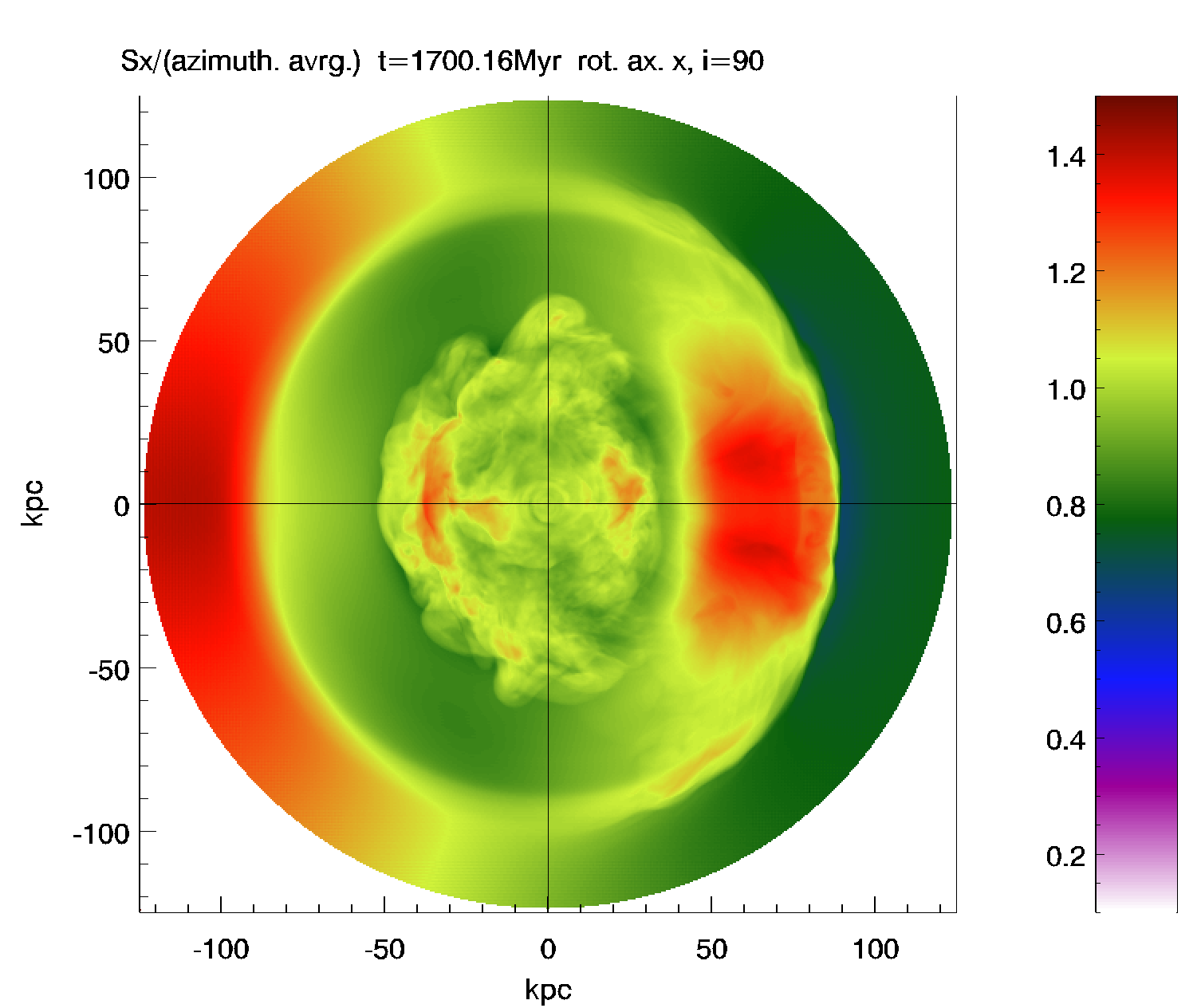}
\newline
\caption{Mock brightness residual maps for different lines-of-sight (LOS). The mock X-ray images are divided by their azimuthally averaged image.   The left column shows the low viscosity case, the right column the high viscosity case. The top row is for the LOS perpendicular to the merger plane. In the subsequent rows the angle between  the LOS and the normal to the orbital plane is 30, 60, and 90 degree. The rotation is around the vertical, i.e.~N-S axis. For rotation around the E-W axis see Fig.~\ref{fig:angle_res_EW}. 
\newline
At high viscosity, the outer CF appears as a smooth arc with a sharp edge, whereas the CF appears ragged and fuzzy at low viscosity. While this statement is true for all LOSs, the overall morphology of the observed brightness residuals in Virgo agree with the top two rows only (compare to Fig.~1 in \citealt{Roediger2011}). }
\label{fig:angle_res_NS}
\end{figure*}
\begin{figure*}
\hspace{2cm} $f_{\mu}=10^{-3}$ \hfill $f_{\mu}=0.1$ \hfill\phantom{x}\newline
\rotatebox{90}{$i=0$, LOS $\perp$ to orbital plane}
\includegraphics[trim=300 0 380 100,clip,angle=90,width=0.41\textwidth]{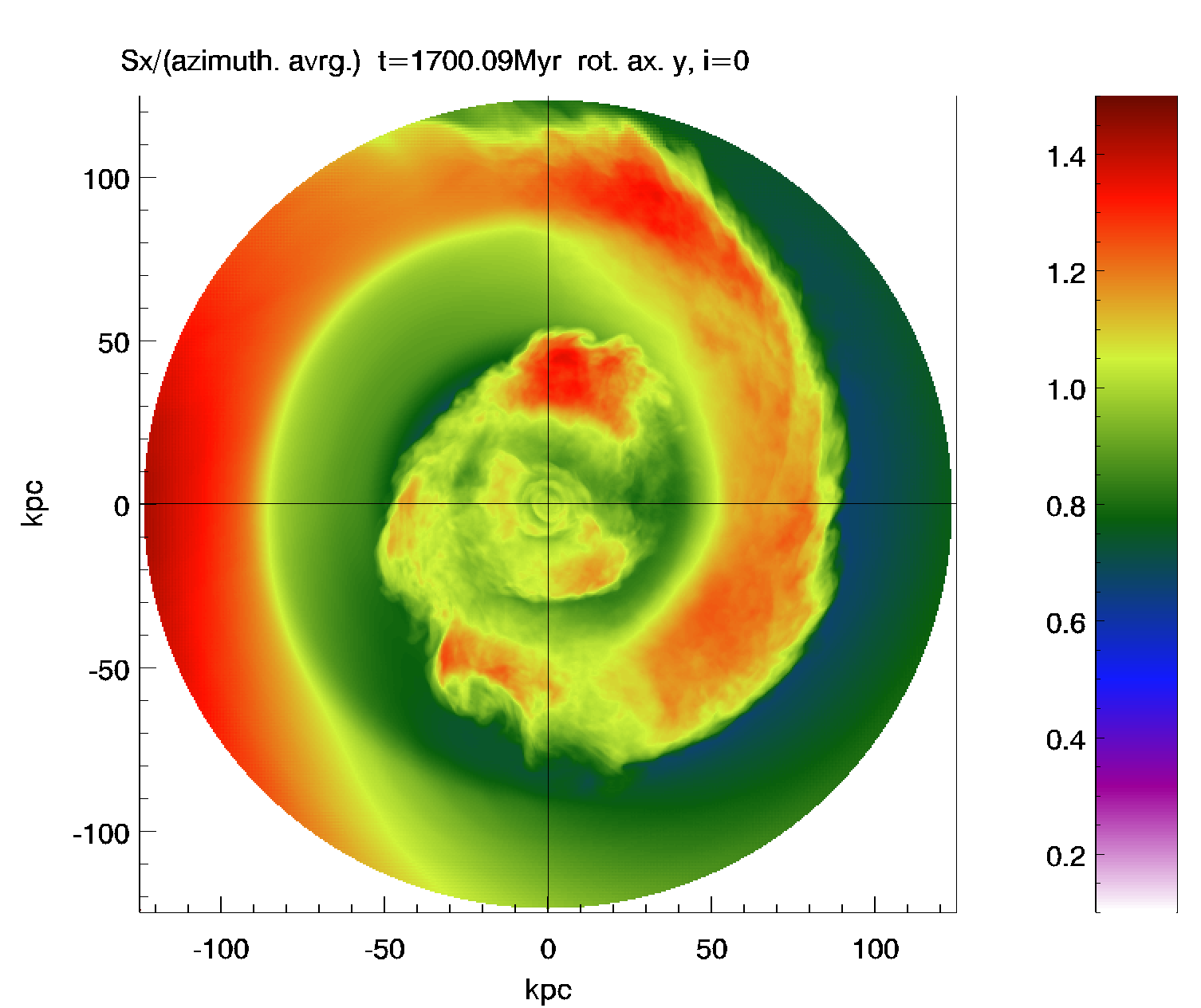}
\includegraphics[trim=300 0 380 100,clip,angle=90,width=0.41\textwidth]{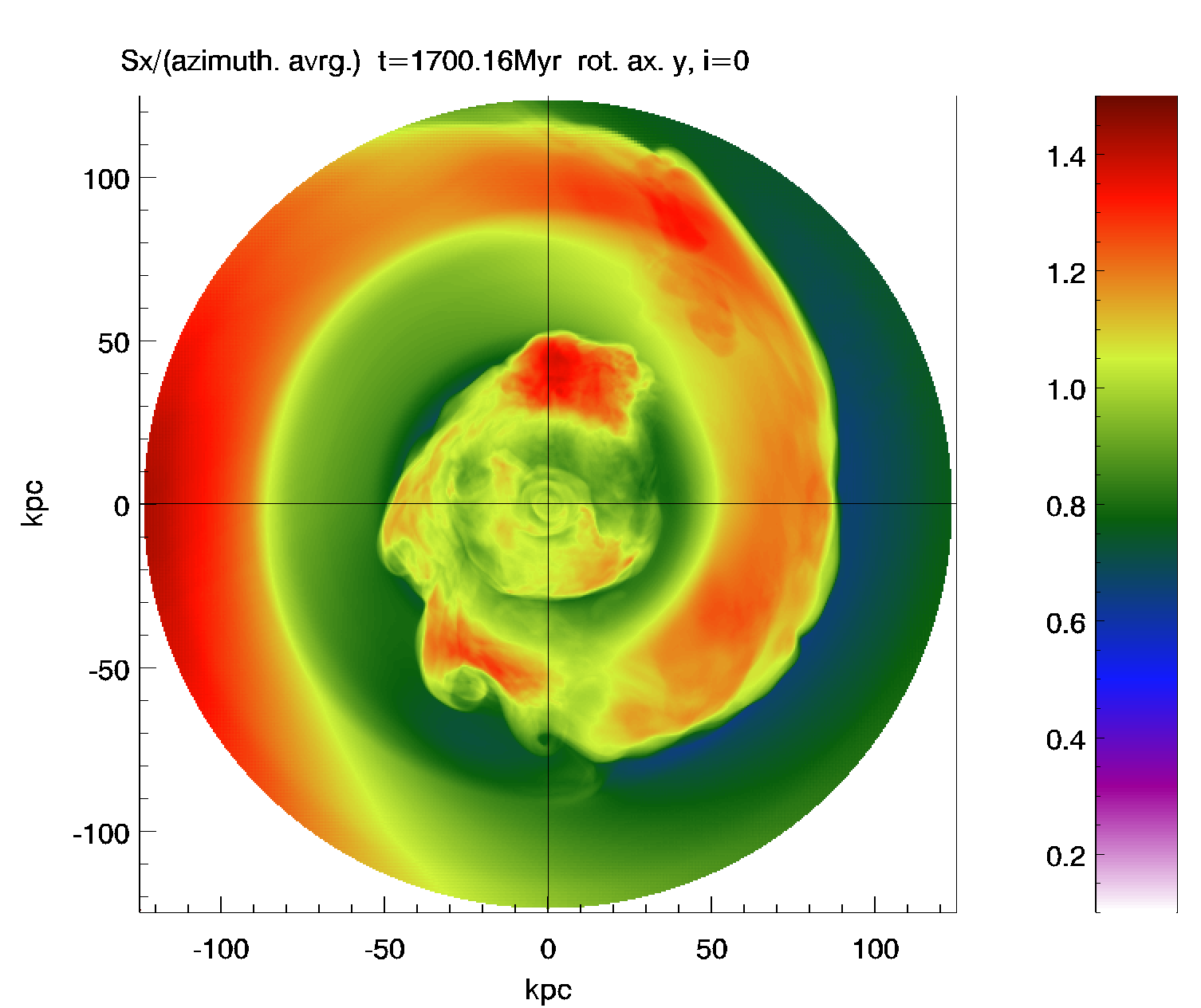}
\includegraphics[trim=1300 100 0 100,clip,angle=0,height=5.5cm]{viscangleResiduals_ax_x_ang_0}
\newline
\rotatebox{90}{\hspace{2cm}$i=30$}
\includegraphics[trim=300 0 380 100,clip,angle=90,width=0.41\textwidth]{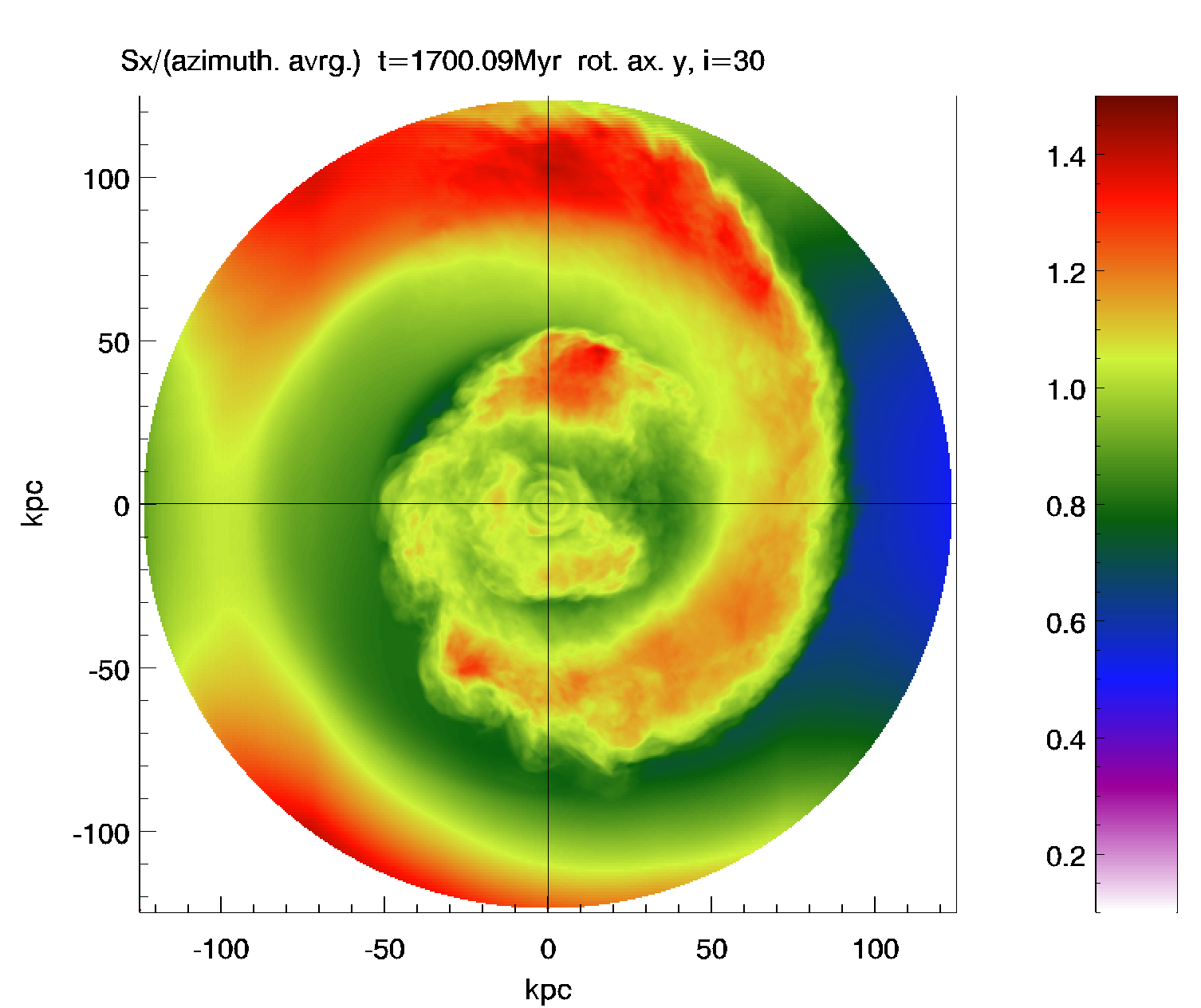}
\includegraphics[trim=300 0 380 100,clip,angle=90,width=0.41\textwidth]{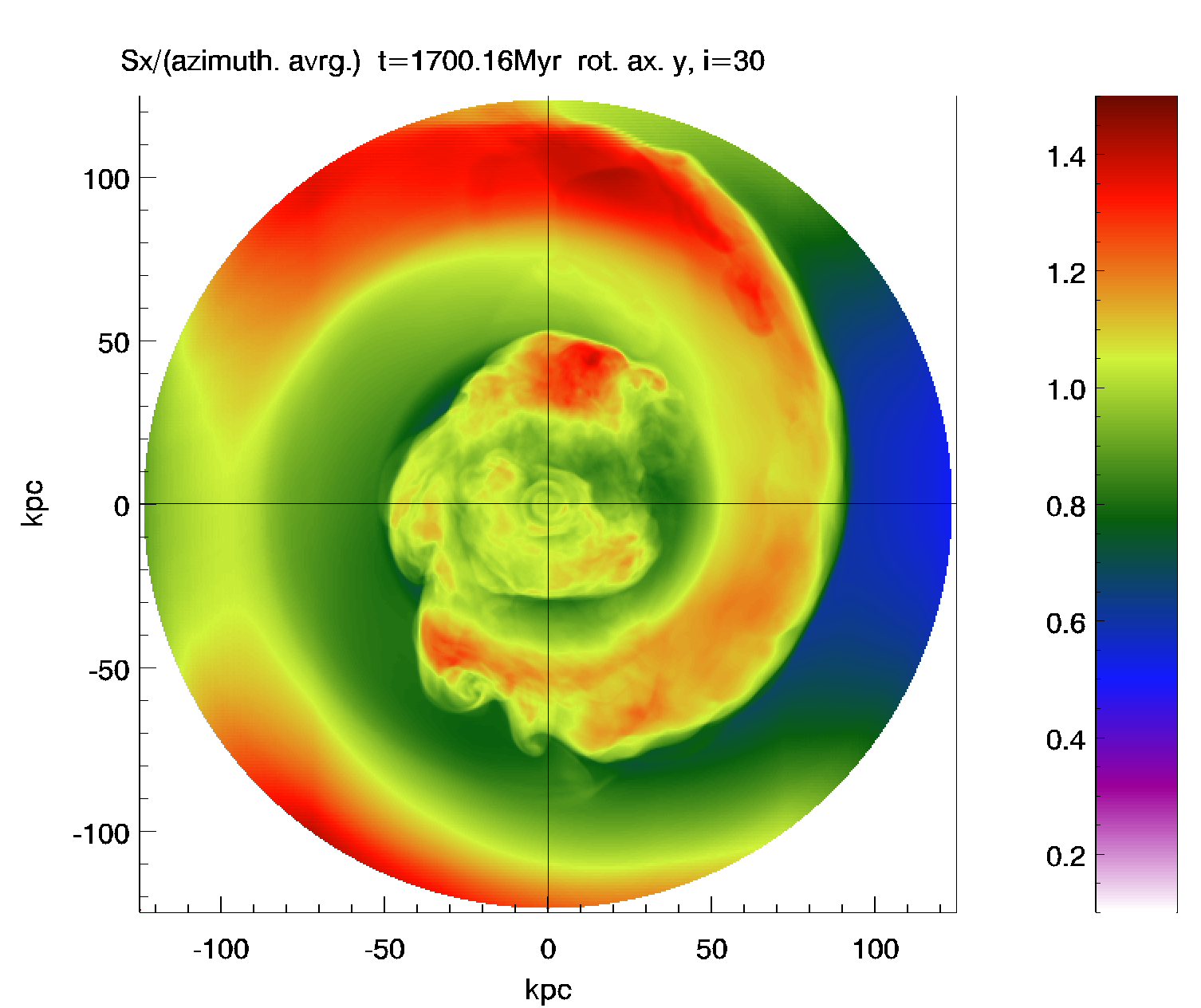}
\newline
\rotatebox{90}{\hspace{2cm}$i=60$}
\includegraphics[trim=300 0 380 100,clip,angle=90,width=0.41\textwidth]{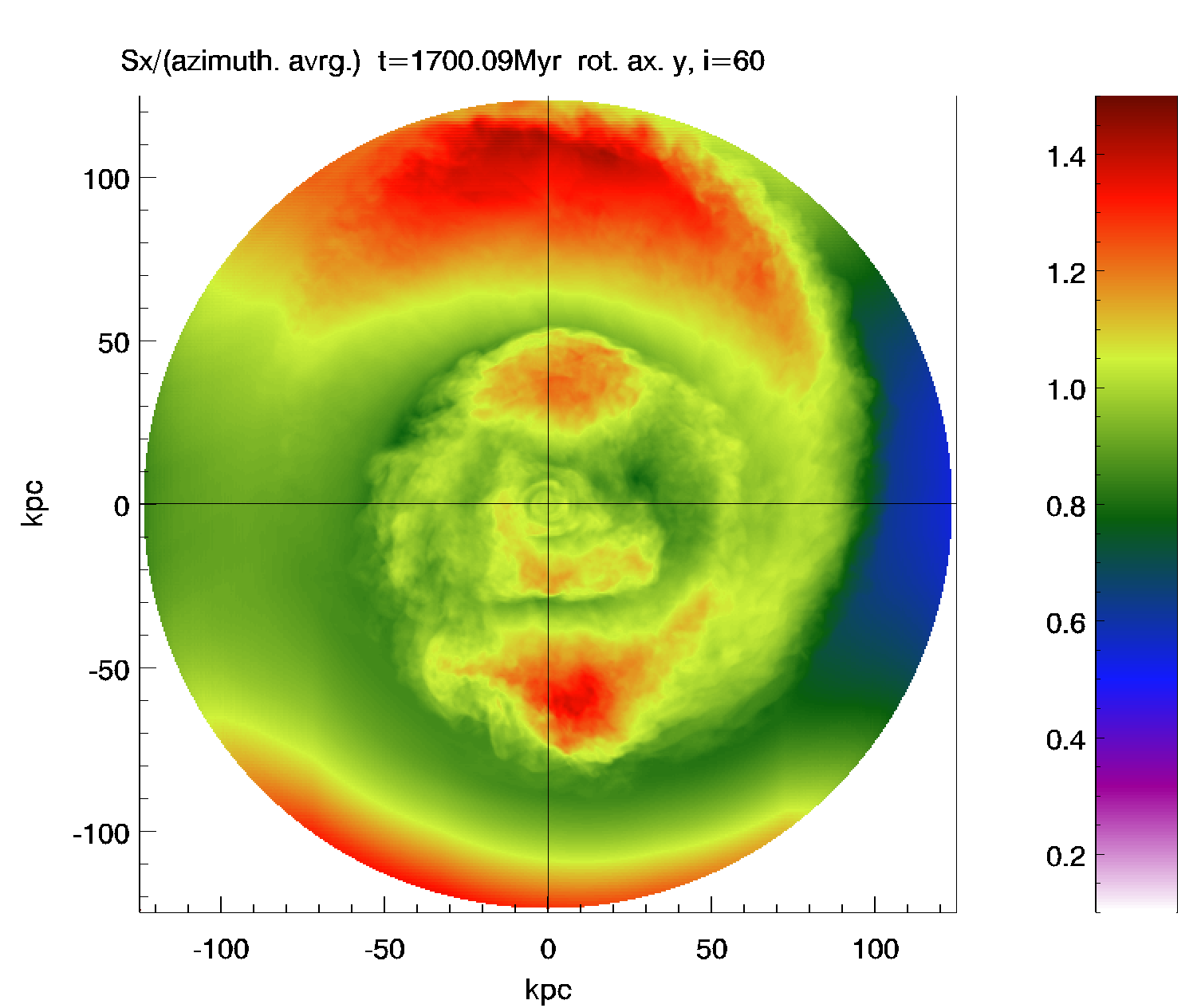}
\includegraphics[trim=300 0 380 100,clip,angle=90,width=0.41\textwidth]{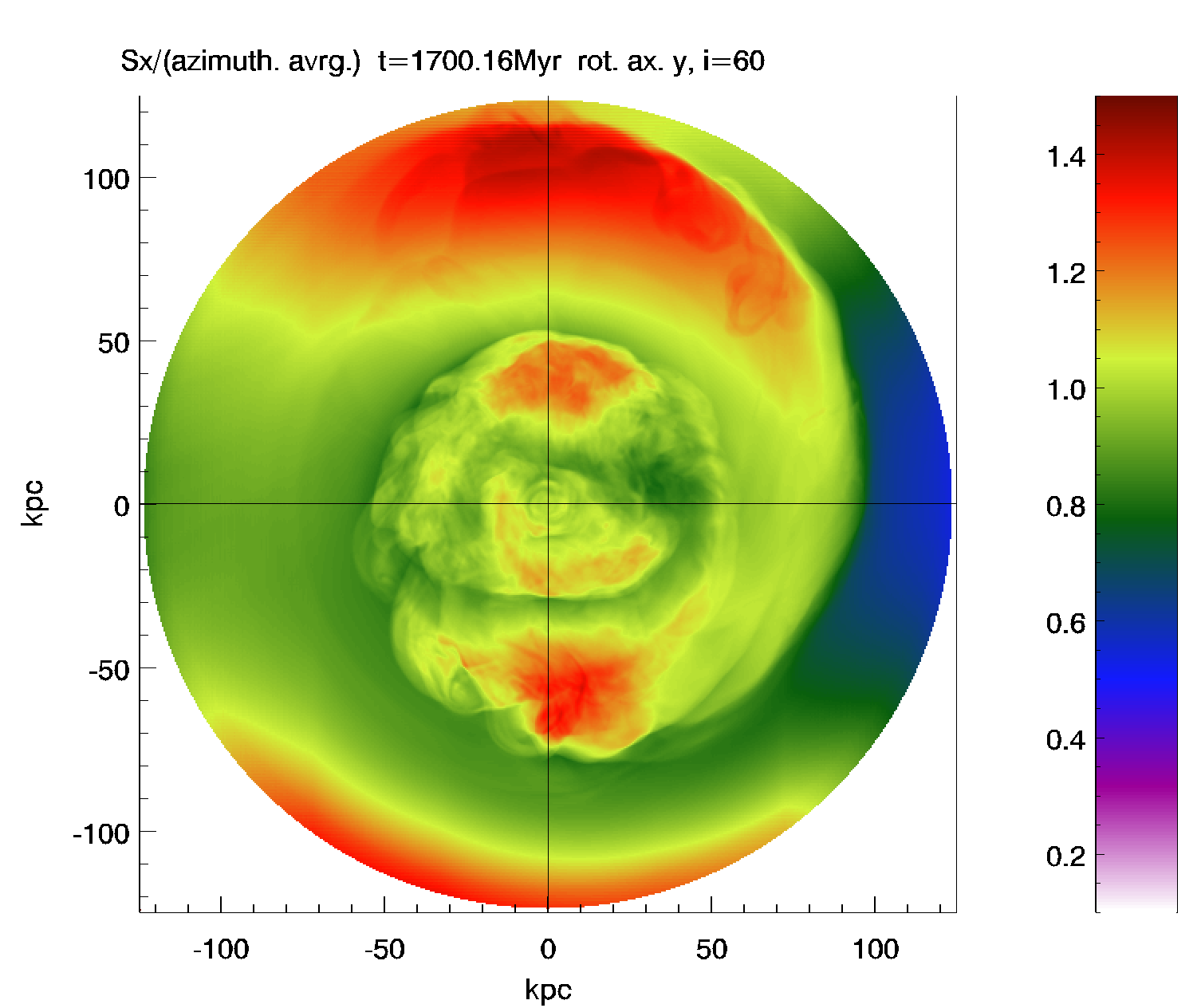}
\newline
\rotatebox{90}{$i=90$, LOS in orbital plane}
\includegraphics[trim=300 0 380 100,clip,angle=90,width=0.41\textwidth]{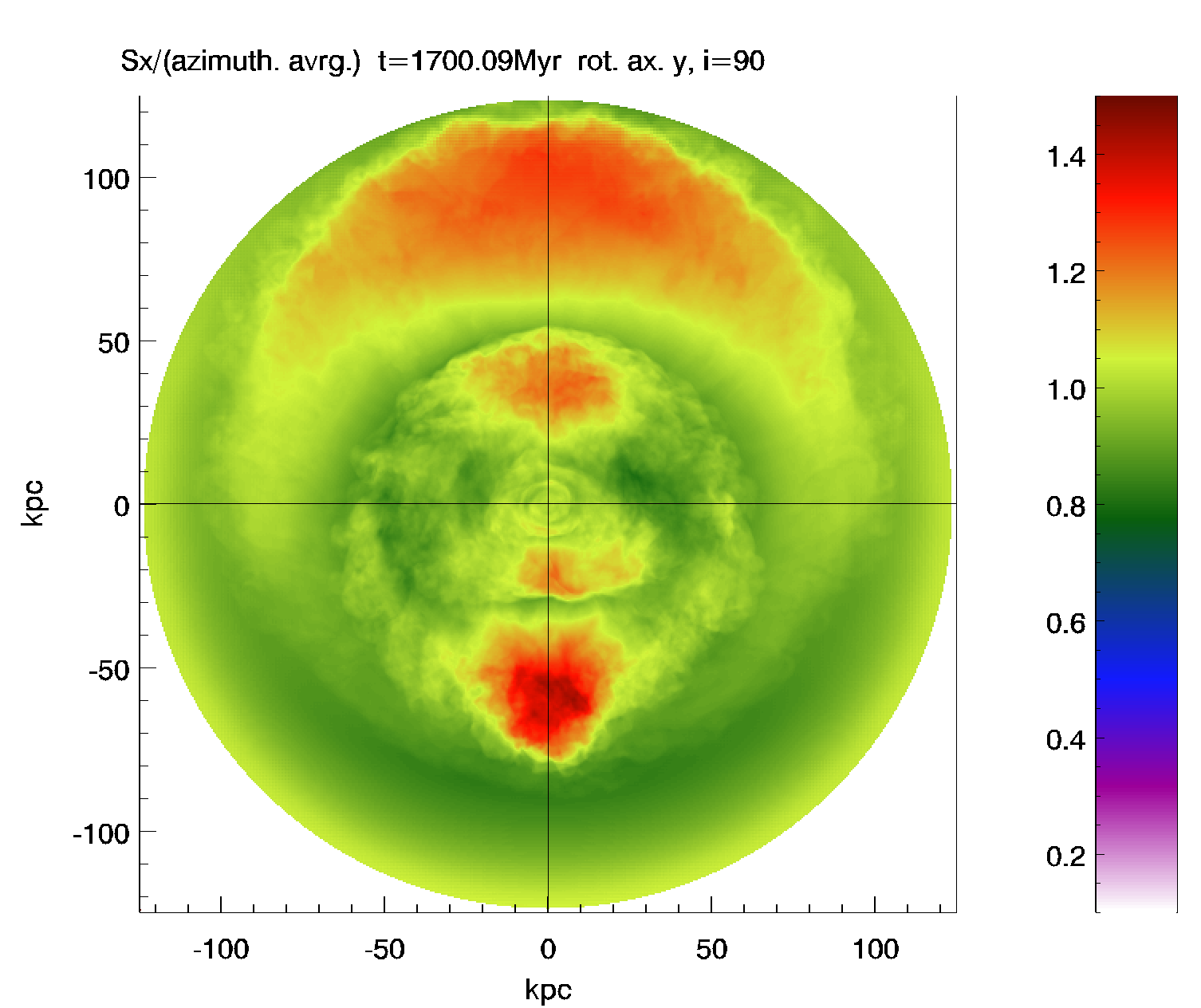}
\includegraphics[trim=300 0 380 100,clip,angle=90,width=0.41\textwidth]{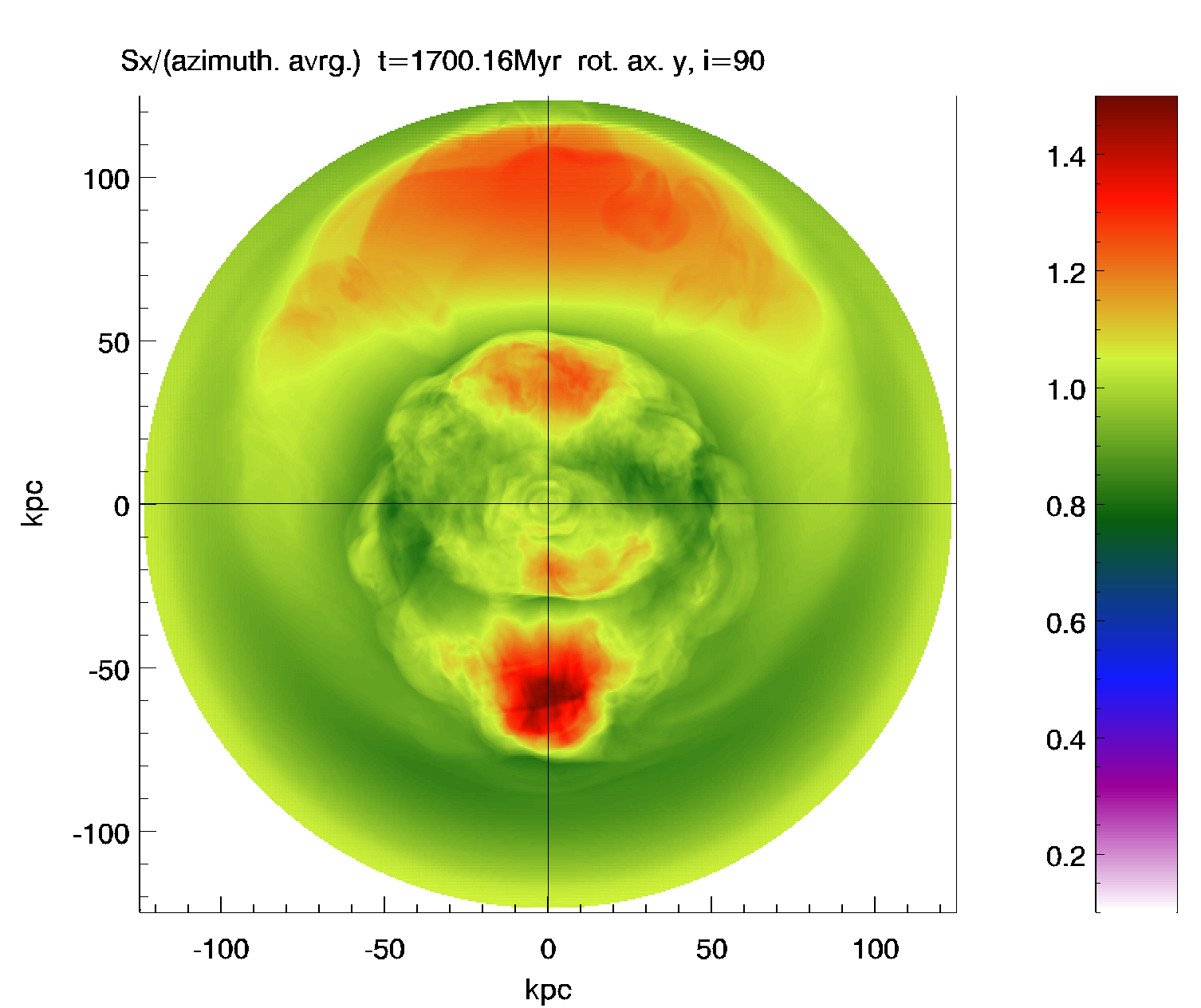}
\newline
\caption{Same as Fig.~\ref{fig:angle_res_NS}, but rotation around the horizontal, i.e.~E-W axis.}
\label{fig:angle_res_EW}
\end{figure*}
%
For our analysis we have assumed a favorable LOS, i.e.~perpendicular to the orbital plane. The spiral-like appearance of the Virgo CFs favors a LOS close to this configuration. To test the dependence of the reported features on the merger geometry we have produced synthetic X-ray images for different LOSs. To enhance the relevant features, we have converted the X-ray images into brightness residual maps by dividing each image by its azimuthal average. The resulting residual maps are shown in Figs.~\ref{fig:angle_res_NS} and \ref{fig:angle_res_EW}. The high and low viscosity case differ clearly for all LOSs. At high viscosity, the outer CF always appears as a smooth arc with a sharp edge, whereas the edge is blurred out by the KHI at low viscosity, and the CF appears ragged and fuzzy.  However, if the merger is close to the plane of the sky, the front in the NW is sharp, even at low viscosity. The fact that the observed northern front in Virgo shows a rather sharp edge in the NW supports the conclusion that the merger in Virgo occurred almost in the plane of the sky, as suggested by the morphology of the brightness residual maps. In this case, the best location to study the difference between existing and suppressed KHIs will be the front between N and NE, not in the NW.

\subsubsection{Resolution and seeding}
At zero viscosity and with ever increasing resolution, the simulations should show progressively smaller KHIs. Naively, one might expect the KHIs to fully smooth out the CF. However, the  presence of KH rolls at a given timestep depends  on the seeds that have been available as well. A full parameter study on this subject is beyond the scope of this paper, but we stress that e.g.~in clouds in the Earth's atmosphere, KH rolls of different wave lengths can be observed simultaneously without the cloud boundary being fully blurred out. In our simulations we simultaneously observe KH rolls of length scales differing by a factor of a few as well.  Hence we conclude that individual KH rolls may be observable with sufficient spatial resolution. Moreover, a blurring out of the CF edge should be clearly detectable in the brightness profiles as shown above. 

\subsubsection{Simplification of viscosity}

Our simulations  describe the momentum diffusion in the ICM with an \emph{effective} viscosity that includes the expected Spitzer temperature dependence. Nature may be more complex than this. The effective ICM viscosity on scales of $\gtrsim 0.1\Kpc$ arises from a complicated interplay of the plasma with the magnetic field. Various processes, e.g.~the gas sloshing itself, impact the amplitude and geometry of the magnetic fields in the cluster core, and the effective viscosity could vary spatially and temporally. Furthermore, due to the tiny electron and  proton gyro-radii, the transport of heat and momentum perpendicular to the magnetic field lines is expected  to be strongly, but not  necessarily fully, suppressed.  This should lead to anisotropic heat and momentum transport preferentially along the field lines. We discuss the effect of magnetic fields and thermal conduction on the KHIs at the CF in the next subsection, here we discuss the implications of a possibly more complex nature of  viscosity.

In the anisotropic case, perturbations perpendicular to the magnetic field lines can be unstable. The instabilities are only suppressed along the field lines. In a tangled field geometry, however, the suppression of the instabilities is mostly isotropic. \citet{Dong2009} demonstrated these effects for buoyant cavities in the ICM. Thus, the effective viscosity relevant for the sloshing CFs will depend on the field geometry. Simulations of MHD sloshing (\citealt{ZuHone2011}) demonstrated that sloshing does not fully erase the field tangling. Moreover, the ordered component of the magnetic fields would be preferentially aligned with the CFs. Given the merger geometry in Virgo, we expect the ordered magnetic fields mostly in the plane of the sky. Consequently, we would observe mostly the KHI modes along the magnetic field lines, and our conclusions regarding the ability of the viscosity to suppress the observable KHIs should remain valid.

The variations in local magnetic field structure could  vary the local effective viscosity, and we need to consider to what extent the effective viscosity inferred at the CFs is representative for the ICM as a whole. The simulations of \citet{ZuHone2011} indicate that the field amplification is not restricted to a narrow layer close to the fronts, but seems to extend throughout the core region. Therefore the effective viscosity inferred from the CF structure may be representative for the entire cluster core region. 
The relation of the viscosity inferred at the CF to the viscosity of the ICM well outside the sloshing region may differ, however.

\subsection{Alternative processes suppressing the KHI}
Apart from an effective viscosity,  aligned magnetic fields or thermal conduction may lead to a suppression of the KHI at CFs. We discuss to what extent these processes can be distinguished observationally.

\subsubsection{Magnetic fields aligned with the cold fronts}

Sufficiently strong magnetic fields parallel to the shear flow interface suppress the KHI (\citealt{Chandrasekhar1961}). Strongly magnetized magnetic draping layers have been shown to form along merger cold fronts, i.e.~along the leading edges of subclusters or galaxies moving through the magnetized ICM of their host clusters (\citealt{Lyutikov2006,Dursi2008}, see \citealt{Vikhlinin2002} for an application to the merger CF in Abell 3667). Such merger CFs are the contact discontinuities between the cooler atmosphere of the subcluster or galaxy and the hotter ICM of the host cluster. 
 
At first glance one might expect that a similar magnetic draping layer forms at sloshing cold fronts as well, and could suppress the KHI. However, the mechanism that forms sloshing CFs is fundamentally different from the one responsible for the merger CFs. It is not related to a body moving through an ambient gas, but it is a nonlinear wave phenomenon of the ICM in the cluster potential (Nulsen \& Roediger, in prep.). As a consequence, gas sloshing leads to magnetic field patterns significantly different from the draping layers at the merger CFs. Sloshing of a magnetized ICM leads to an amplified and ordered magnetic field on the \emph{inside, the colder side}, of the sloshing CFs, not on the outer and hotter side (see MHD simulations of \citealt{ZuHone2011}). The amplification and ``combing" of the field is not restricted to thin layers near the front, but  layers of amplified fields can be found at the CFs and elsewhere on their inside. These layers are not necessarily thin, but can be  tens of kpc wide. The draping layers at the merger CFs finally saturate at equipartition with the thermal pressure, but the sloshing-amplified magnetic field strength depends on the assumed initial field strength. While the average field strength throughout the cluster core has a somewhat weaker dependence on the initial strength, the field strength directly at the CFs is sensitive to the initial strength (\citealt{ZuHone2011}). Consequently, the ability of the amplified field to suppress the growth of the KHI depends on the initial magnetic field strength, and only the strongest plausible initial magnetic field reached amplifications sufficient to suppress the KHI, which is especially true for the outer CFs. 

\citet{ZuHone2011} emphasized that their simulations demonstrated the \emph{ability} of observationally plausible magnetic fields to suppress the KHI at sloshing CFs. We would suggest a somewhat different interpretation of their simulation results: Given that only the strongest plausible fields clearly suppressed the KHI, which is especially true for the outer CFs,  this indicates that observed magnetic field strengths in clusters are in a range where they \emph{may or may not be able to suppress the KHIs} at sloshing CFs. The answer may differ from cluster to cluster, and indeed both smooth and ragged CFs are observed (see Sect.~\ref{sec:compare_others}). A detailed investigation of individual clusters is required, where the actual merger history, and thus dynamical ICM properties, as well as the magnetic field strengths are observationally constrained. 

Observations show both ragged and smooth CFs (see Sect.~\ref{sec:compare_others}). In the case of smooth CFs, it would be helpful if we could distinguish whether the suppression of the KHIs occurred by viscosity or amplified magnetic fields. If a thin layer of magnetic fields aligned with the CF suppresses the KHI, the field strength in this layer needs to exceed about 10\% of the thermal pressure in this layer (\citealt{Vikhlinin2002}), which results in an approximately 20\%  surface brightness depression, which could be detected in deep observations. However, the sloshing-amplified fields do not have a simple draping layer geometry, but the amplified layers may be rather wide and extend far below the CFs. Consequently, the resulting reductions in thermal ICM pressure would not be localized at the CF, but the reduction would be gradual from the CF inwards, and it may be challenging to recognize the  presence of this gradual reduction  in thermal ICM pressure profiles. However, the simulations of \citet{ZuHone2011} suggest that sloshing forms multiple layers, streaks of amplified magnetic fields below the CFs. Their magnetic pressure  leads to corresponding streaks of reduced ICM density that should be detectable as ripples in surface brightness. It will be interesting to compare not only the structure \emph{at} the CFs, but also \emph{below} the CFs in order to distinguish the cases of high viscosity and amplified aligned magnetic fields.
Additionally, a significantly amplified field over the whole cluster core  may well give rise to radio synchrotron emission and thus radio minihalos, as suggested by \citet{Mazzotta2008} and shown by \citet{ZuHone2011}. It will be interesting to investigate the correlation between the presence of radio minihalos and the smoothness of sloshing CFs. 
  
\subsubsection{Magnetic fields and anisotropic viscosity}

It will  be interesting to study the combined effect of magnetic fields and (anisotropic) viscosity, because they have opposite effects and mutual dependencies. In the standard picture, zero magnetic fields imply a high viscosity, and thus sharp CFs that form smooth arcs. Increasing the magnetic field strength is expected to reduce the effective viscosity on the one hand, thus permitting KHIs larger than a certain wavelength. Thus, as a first effect, the presence of magnetic fields facilitates the growth of KHIs.  This could be observed as a ragged structure of the CF arc, and the density and brightness jump being washed out. The instabilities create turbulence and may tangle and amplify the fields,  reducing the effective viscosity further. However, once the magnetic fields are strong enough, they may, themselves, suppress the growth of the KHI. Their effect on the ICM density and temperature structure should be very similar to the viscous suppression of the KHI, but the enhanced magnetic fields may be revealed by the radio minihalos. 

\subsubsection{Thermal conduction}

\citet{Xiang2007} discuss the impact of thermal conduction on merger CFs. Given that the formation mechanism of merger and sloshing CFs differs, the direct application of their results to sloshing CFs is unclear. \citet{ZuHone2012} presented MHD simulations of gas sloshing and included an anisotropic thermal conduction, where conduction was fully or strongly suppressed perpendicular to the magnetic field lines. The sloshing leads to an approximate alignment of the magnetic fields with the fronts, but the cool regions below the fronts are still connected with hotter regions above the fronts, and full Spitzer conduction along the field lines has a significant impact on the CFs. The temperature and density contrasts are strongly reduced, and the fronts are less sharp, the discontinuity is washed out. The resulting finite width of the interface suppresses KHIs of perturbation length scales about  several times as long as the interface width (\citealt{Chandrasekhar1961}). 

In contrast to viscosity, thermal conduction widens the temperature and density discontinuity, whereas viscosity widens the discontinuity in shear velocity. This leads to a characteristic difference in the CF structure: Thermal conduction should lead to  CFs free of KHI below a certain wavelength, but a washed out discontinuity in density and temperature, i.e.~the edge will not be sharp in either of these two quantities. Viscosity, on the other hand, will  lead to a smooth-arc CF as well, but a sharp one, as neither the density nor the temperature discontinuity are washed out. The \emph{presence} of KHIs  leads to an apparently washed-out CF in surface brightness as well, but in this case, individual KH rolls may be identifiable. Moreover, the cold gas dominates the emission measure in the mixing layer induced by the KHIs, and we expect a steep jump in projected temperature across the front. In contrast, thermal conduction truly washes out the temperature jump, leading to a different profile in projected temperature across the front.

If either thermal conduction or viscosity suppresses the KHI not fully but only below an observable length scale, KH rolls of larger length should be recognizable. As just mentioned, if viscosity suppresses the small KHIs, individual stretches of the CF shorter than the limiting length scale should appear as sharp edges, and the existing, larger KH rolls may be very pronounced (see the big KH rolls in the SW of the simulated Virgo cluster in Fig.~\ref{fig:Xray}). In the case thermal conduction suppresses the small-scale KHIs, the simulations of \citet{ZuHone2012} indicate that  short stretches of the CF should not be sharp but be washed out.

The CF structure resulting from the combined effect of thermal conduction and viscosity has yet to be determined.

\subsection{Comparison to other clusters} \label{sec:compare_others}
The detailed structure of sloshing CFs differs between clusters. In some clusters they appear as smooth arcs within the observational resolution, e.g.~Abell 2142 (\citealt{Markevitch2000,Owers2009hifid}) and Abell 2204 (\citealt{Sanders2009a2204}). The interpretation of the structure of the sloshing CFs in Perseus (\citealt{Fabian2006}) and Centaurus (\citealt{Fabian2005centaurus}) is complicated by interference of AGN activity. 
 However, the CFs in Abell 496 (\citealt{Dupke2007}) have a distinct boxy or even concave appearance that resembles KH rolls (\citealt{Roediger2012a496}). The CF towards the west is a doublet similar to the multiple adjacent edges reported here. The northern CF  seems to curve around in the NE, but becomes washed out in the east.  The sloshing CFs in the group-group merger NGC 7618/UGC 12491 display a triangular-shaped nose and wings reminiscent of  KH rolls shown in Fig.~\ref{fig:Xray} (\citealt{Roediger2012n7618}).  We  note that a significant fraction of the observed cold fronts in cluster cores are not shaped like the long arcs in spiral or staggered arrangements predicted by numerical simulations, but several clusters exhibit only short stretches of clear cold fronts. It will be interesting to revisit the observational data and attempt to distinguish whether this is due to the orientation of the merger with respect to our LOS, an intrinsic ellipticity of the cluster or a washing out of the CFs by instabilities.

\section{Summary} \label{sec:summary}

In this paper we present the first detailed attempt to constrain the effective viscosity of the ICM from the presence or absence of KHIs at sloshing CFs. We focussed on the Virgo cluster for its proximity, brightness and favorable merger geometry. We performed viscous hydrodynamical simulations tailored for the Virgo cluster, compared them to existing \textit{XMM-Newton} observations and made predictions for future deep \textit{Chandra} observations. 
Our key results are: 
\begin{itemize}
\item At the Virgo cluster cold fronts, a viscosity of 10\% of the Spitzer value suppresses the KHI for scales below tens of kpc. In contrast,  the KHI occurs at  $\sim 15\Kpc$ scales at viscosities below 1\% of the Spitzer value (Fig.~\ref{fig:tempslices}). 
\item We identify the outer CF at $\sim 90\Kpc$ to the north and north-east of the Virgo center as the best target for distinguishing different viscosities observationally, because this front is not disturbed by nuclear outbursts of M87, and it is subject to a sufficient shear flow which potentially induces the KHI. 
\item Our simulations predict observationally distinguishable structures at this CF: at viscosities of $\gtrsim10$\% of the Spitzer value, the KHI is suppressed along the northern CF, it appears as a smooth arc (bottom panel of Fig.~\ref{fig:Xray}). Surface brightness profiles across the front are consistent with a sharp density discontinuity (bottom panel of Fig.~\ref{fig:sbp}). At  viscosities $\lesssim 1$\% of the Spitzer value, the front has a ragged, sawtooth-like appearance (top panel of Fig.~\ref{fig:Xray}). Individual KH rolls at $\sim 15 \Kpc$ size can be identified, and the front splits up into multiple, adjacent brightness edges.  Surface brightness profiles across the front are shallower than expected for a density discontinuity, i.e.~the discontinuity appears blurred out. Profiles in narrow wedges show multiple jumps arising from individual KH rolls (top panel of Fig.~\ref{fig:sbp}). 
\item We re-analyze archival \textit{XMM-Newton} data for corresponding signatures. The northern cold front appears to be less sharp in the north-east than in the north-west (Fig.~\ref{fig:cf}), indicating a blurring out of the discontinuity in the north-east by KHIs and thus a low ICM viscosity. Due to low count statistics and the non-uniform detector response these data do not allow the detection of individual KH rolls.  
\item A 300 ks \textit{Chandra} observation of the north-eastern front would confirm the presence or absence of KHIs both in images (Fig.~\ref{fig:sim}) and in profiles across the front (Fig.~\ref{fig:sbp}) and thus constrain the effective ICM viscosity. 
\item We discuss in detail the differences in KHI suppression by viscosity, magnetic fields and thermal conduction, and how these processes can be distinguished observationally.
\end{itemize}

\section*{Acknowledgments}
E.R.~acknowledges support by the Priority Programmes 1177 ("Witnesses of Cosmic History") and 1573 ("Physics of the Interstellar Medium") of the DFG (German Research Foundation),  the supercomputing grants NIC 4368 and 5027 at the J\"ulich Supercomputing Center, a visiting scientist fellowship of the Smithsonian Astrophysical Observatory, and the hospitality of the Center for Astrophysics in Cambridge. We thank Marcus Br\"uggen for helpful discussions and the referee for his/her useful suggestions.

This work was supported by NASA grant NAS8-03060.

%

\begin{thebibliography}{54}
\expandafter\ifx\csname natexlab\endcsname\relax\def\natexlab#1{#1}\fi

\bibitem[{Ascasibar \& Markevitch(2006)}]{Ascasibar2006}
Ascasibar, Y., \& Markevitch, M. 2006, ApJ, 650, 102

\bibitem[{Balbus \& Reynolds(2010)}]{Balbus2010a}
Balbus, S.~A., \& Reynolds, C.~S. 2010, ApJ, 720, L97

\bibitem[{Chandrasekhar(1961)}]{Chandrasekhar1961}
Chandrasekhar, S. 1961, {Hydrodynamic and hydromagnetic stability} (Oxford:
  Clarendon)

\bibitem[{Churazov \& Inogamov(2004)}]{Churazov2004}
Churazov, E., \& Inogamov, N. 2004, MNRAS, 350, L52

\bibitem[{Churazov {et~al.}(2012)Churazov, Vikhlinin, Zhuravleva, Schekochihin,
  Parrish, Sunyaev, Forman, B\"{o}hringer, \& Randall}]{Churazov2012}
Churazov, E., Vikhlinin, A., Zhuravleva, I., {et~al.} 2012, MNRAS, 421, 1123

\bibitem[{Dong \& Stone(2009)}]{Dong2009}
Dong, R., \& Stone, J.~M. 2009, ApJ, 704, 1309

\bibitem[{Dubey {et~al.}(2009)Dubey, Antypas, Ganapathy, Reid, Riley, Sheeler,
  Siegel, \& Weide}]{Dubey2009}
Dubey, A., Antypas, K., Ganapathy, M.~K., {et~al.} 2009, Parallel Computing,
  35, 512

\bibitem[{Dupke {et~al.}(2007)Dupke, {White III}, \& Bregman}]{Dupke2007}
Dupke, R., {White III}, R.~E., \& Bregman, J.~N. 2007, ApJ, 671, 181

\bibitem[{Dursi \& Pfrommer(2008)}]{Dursi2008}
Dursi, L.~J., \& Pfrommer, C. 2008, ApJ, 677, 993

\bibitem[{Fabian {et~al.}(2005{\natexlab{a}})Fabian, Reynolds, Taylor, \&
  Dunn}]{Fabian2005}
Fabian, A.~C., Reynolds, C.~S., Taylor, G.~B., \& Dunn, R. J.~H.
  2005{\natexlab{a}}, MNRAS, 363, 891

\bibitem[{Fabian {et~al.}(2003)Fabian, Sanders, Crawford, Conselice, Gallagher,
  \& Wyse}]{Fabian2003halpha}
Fabian, A.~C., Sanders, J.~S., Crawford, C.~S., {et~al.} 2003, MNRAS, 344, L48

\bibitem[{Fabian {et~al.}(2005{\natexlab{b}})Fabian, Sanders, Taylor, \&
  Allen}]{Fabian2005centaurus}
Fabian, A.~C., Sanders, J.~S., Taylor, G.~B., \& Allen, S.~W.
  2005{\natexlab{b}}, MNRAS, 360, L20

\bibitem[{Fabian {et~al.}(2006)Fabian, Sanders, Taylor, Allen, Crawford,
  Johnstone, \& Iwasawa}]{Fabian2006}
Fabian, A.~C., Sanders, J.~S., Taylor, G.~B., {et~al.} 2006, MNRAS, 366, 417

\bibitem[{Forman {et~al.}(2007)Forman, Jones, Churazov, Markevitch, Nulsen,
  Vikhlinin, Begelman, Bohringer, Eilek, Heinz, Kraft, Owen, \&
  Pahre}]{Forman2007}
Forman, W.~R., Jones, C., Churazov, E., {et~al.} 2007, ApJ, 665, 1057

\bibitem[{Ghizzardi {et~al.}(2010)Ghizzardi, Rossetti, \&
  Molendi}]{Ghizzardi2010}
Ghizzardi, S., Rossetti, M., \& Molendi, S. 2010, A\&A, 516, A32

\bibitem[{Junk {et~al.}(2010)Junk, Walch, Heitsch, Burkert, Wetzstein,
  Schartmann, \& Price}]{Junk2010}
Junk, V., Walch, S., Heitsch, F., {et~al.} 2010, MNRAS, 407, 1933

\bibitem[{Kaiser {et~al.}(2005)Kaiser, Pavlovski, Pope, \&
  Fangohr}]{Kaiser2005}
Kaiser, C.~R., Pavlovski, G., Pope, E. C.~D., \& Fangohr, H. 2005, MNRAS, 359,
  493

\bibitem[{Keshet {et~al.}(2010)Keshet, Markevitch, Birnboim, \&
  Loeb}]{Keshet2010}
Keshet, U., Markevitch, M., Birnboim, Y., \& Loeb, A. 2010, ApJ, 719, L74

\bibitem[{Kunz(2011)}]{Kunz2011}
Kunz, M.~W. 2011, MNRAS, 417, 602

\bibitem[{Kunz {et~al.}(2012)Kunz, Bogdanovi\'{c}, Reynolds, \&
  Stone}]{Kunz2012}
Kunz, M.~W., Bogdanovi\'{c}, T., Reynolds, C.~S., \& Stone, J.~M. 2012, ApJ,
  754, 122

\bibitem[{Lyutikov(2006)}]{Lyutikov2006}
Lyutikov, M. 2006, MNRAS, 373, 73

\bibitem[{Markevitch \& Vikhlinin(2007)}]{Markevitch2007}
Markevitch, M., \& Vikhlinin, A. 2007, Physics Reports, 443, 1

\bibitem[{Markevitch {et~al.}(2003)Markevitch, Vikhlinin, \&
  Forman}]{Markevitch2003}
Markevitch, M., Vikhlinin, A., \& Forman, W.~R. 2003, in ASP Conference
  Proceedings, Vol. 301, Matter and Energy in Clusters of Galaxies, ed.
  S.~Bowyer \& C.-Y. Hwang (San Francisco: Astron. Soc. Pac.), 37

\bibitem[{Markevitch {et~al.}(2001)Markevitch, Vikhlinin, \&
  Mazzotta}]{Markevitch2001}
Markevitch, M., Vikhlinin, A., \& Mazzotta, P. 2001, ApJ, 562, L153

\bibitem[{Markevitch {et~al.}(2000)Markevitch, Ponman, Nulsen, Bautz, Burke,
  David, Davis, Donnelly, Forman, Jones, Kaastra, Kellogg, Kim, Kolodziejczak,
  Mazzotta, Pagliaro, Patel, {Van Speybroeck}, Vikhlinin, Vrtilek, Wise, \&
  Zhao}]{Markevitch2000}
Markevitch, M., Ponman, T.~J., Nulsen, P. E.~J., {et~al.} 2000, ApJ, 541, 542

\bibitem[{Mazzotta \& Giacintucci(2008)}]{Mazzotta2008}
Mazzotta, P., \& Giacintucci, S. 2008, ApJ, 675, L9

\bibitem[{Narayan \& Medvedev(2001)}]{Narayan2001}
Narayan, R., \& Medvedev, M.~V. 2001, ApJ, 562, L129

\bibitem[{Nulsen(1982)}]{Nulsen1982}
Nulsen, P. E.~J. 1982, MNRAS, 198, 1007

\bibitem[{Owers {et~al.}(2009)Owers, Nulsen, Couch, \&
  Markevitch}]{Owers2009hifid}
Owers, M.~S., Nulsen, P. E.~J., Couch, W.~J., \& Markevitch, M. 2009, ApJ, 704,
  1349

\bibitem[{Parrish {et~al.}(2010)Parrish, Quataert, \& Sharma}]{Parrish2010}
Parrish, I.~J., Quataert, E., \& Sharma, P. 2010, ApJ, 712, L194

\bibitem[{Rebusco {et~al.}(2005)Rebusco, Churazov, B\"{o}hringer, \&
  Forman}]{Rebusco2005}
Rebusco, P., Churazov, E., B\"{o}hringer, H., \& Forman, W.~R. 2005, MNRAS,
  359, 1041

\bibitem[{Rebusco {et~al.}(2006)Rebusco, Churazov, B\"{o}hringer, \&
  Forman}]{Rebusco2006}
---. 2006, MNRAS, 372, 1840

\bibitem[{Reynolds {et~al.}(2005)Reynolds, McKernan, Fabian, Stone, \&
  Vernaleo}]{Reynolds2005}
Reynolds, C.~S., McKernan, B., Fabian, A.~C., Stone, J.~M., \& Vernaleo, J.~C.
  2005, MNRAS, 357, 242

\bibitem[{Roediger \& Br\"{u}ggen(2008)}]{Roediger2008}
Roediger, E., \& Br\"{u}ggen, M. 2008, MNRAS, 388, 465

\bibitem[{Roediger {et~al.}(2007)Roediger, Br\"{u}ggen, Rebusco, B\"{o}hringer,
  \& Churazov}]{Roediger2007bubbles}
Roediger, E., Br\"{u}ggen, M., Rebusco, P., B\"{o}hringer, H., \& Churazov, E.
  2007, MNRAS, 375, 15

\bibitem[{Roediger {et~al.}(2011)Roediger, Br\"{u}ggen, Simionescu,
  B\"{o}hringer, Churazov, \& Forman}]{Roediger2011}
Roediger, E., Br\"{u}ggen, M., Simionescu, A., {et~al.} 2011, MNRAS, 413, 2057

\bibitem[{Roediger {et~al.}(2012{\natexlab{a}})Roediger, Kraft, Machacek,
  Forman, Nulsen, Jones, \& Murray}]{Roediger2012n7618}
Roediger, E., Kraft, R., Machacek, M., {et~al.} 2012{\natexlab{a}}, ApJ, 754,
  147

\bibitem[{Roediger {et~al.}(2012{\natexlab{b}})Roediger, Lovisari, Dupke,
  Ghizzardi, Br\"{u}ggen, Kraft, \& Machacek}]{Roediger2012a496}
Roediger, E., Lovisari, L., Dupke, R., {et~al.} 2012{\natexlab{b}}, MNRAS, 420,
  3632

\bibitem[{Ryu {et~al.}(2011)Ryu, Schleicher, Treumann, Tsagas, \&
  Widrow}]{Ryu2011}
Ryu, D., Schleicher, D. R.~G., Treumann, R.~A., Tsagas, C.~G., \& Widrow, L.~M.
  2011, Space Sci. Rev.

\bibitem[{Sanders {et~al.}(2010)Sanders, Fabian, \& Smith}]{Sanders2010a}
Sanders, J.~S., Fabian, A.~C., \& Smith, R.~K. 2010, MNRAS, 410, no

\bibitem[{Sanders {et~al.}(2009)Sanders, Fabian, \& Taylor}]{Sanders2009a2204}
Sanders, J.~S., Fabian, A.~C., \& Taylor, G.~B. 2009, MNRAS, 393, 71

\bibitem[{Sarazin(1988)}]{Sarazin1988}
Sarazin, C.~L. 1988, Cambridge Astrophysics Series

\bibitem[{Schuecker {et~al.}(2004)Schuecker, Finoguenov, Miniati,
  B\"{o}hringer, \& Briel}]{Schuecker2004}
Schuecker, P., Finoguenov, A., Miniati, F., B\"{o}hringer, H., \& Briel, U.~G.
  2004, A\&A, 426, 387

\bibitem[{Simionescu {et~al.}(2010)Simionescu, Werner, Forman, Miller, Takei,
  B\"{o}hringer, Churazov, \& Nulsen}]{Simionescu2010}
Simionescu, A., Werner, N., Forman, W.~R., {et~al.} 2010, MNRAS, 405, 91

\bibitem[{Spitzer(1956)}]{Spitzer1956}
Spitzer, L. 1956, {Physics of Fully Ionized Gases} (New York: Interscience
  Publishers)

\bibitem[{Sutherland \& Dopita(1993)}]{Sutherland1993}
Sutherland, R.~S., \& Dopita, M.~A. 1993, ApJSS, 88, 253

\bibitem[{Vazza {et~al.}(2012)Vazza, Roediger, \& Br\"{u}ggen}]{Vazza2012}
Vazza, F., Roediger, E., \& Br\"{u}ggen, M. 2012, A\&A, 544, A103

\bibitem[{Vikhlinin {et~al.}(2001)Vikhlinin, Markevitch, \&
  Murray}]{Vikhlinin2001}
Vikhlinin, A., Markevitch, M., \& Murray, S.~S. 2001, ApJ, 551, 160

\bibitem[{Vikhlinin \& Markevitch(2002)}]{Vikhlinin2002}
Vikhlinin, A.~A., \& Markevitch, M. 2002, Astronomy Letters, 28, 495

\bibitem[{Xiang {et~al.}(2007)Xiang, Churazov, Dolag, Springel, \&
  Vikhlinin}]{Xiang2007}
Xiang, F., Churazov, E., Dolag, K., Springel, V., \& Vikhlinin, A. 2007, MNRAS,
  379, 1325

\bibitem[{Zhuravleva {et~al.}(2012)Zhuravleva, Churazov, Kravtsov, \&
  Sunyaev}]{Zhuravleva2012}
Zhuravleva, I., Churazov, E., Kravtsov, A., \& Sunyaev, R. 2012, MNRAS, 422,
  2712

\bibitem[{ZuHone {et~al.}(2010)ZuHone, Markevitch, \& Johnson}]{ZuHone2010}
ZuHone, J.~A., Markevitch, M., \& Johnson, R.~E. 2010, ApJ, 717, 908

\bibitem[{ZuHone {et~al.}(2011)ZuHone, Markevitch, \& Lee}]{ZuHone2011}
ZuHone, J.~A., Markevitch, M., \& Lee, D. 2011, ApJ, 743, 16

\bibitem[{ZuHone {et~al.}(2012)ZuHone, Markevitch, Ruszkowski, \&
  Lee}]{ZuHone2012}
ZuHone, J.~A., Markevitch, M., Ruszkowski, M., \& Lee, D. 2012, eprint
  arXiv:1204.6005

\end{thebibliography}

\appendix

\section{Kelvin-Helmholtz instability at low resolution} \label{sec:appendix}

\begin{figure*}
\includegraphics[width=\textwidth]{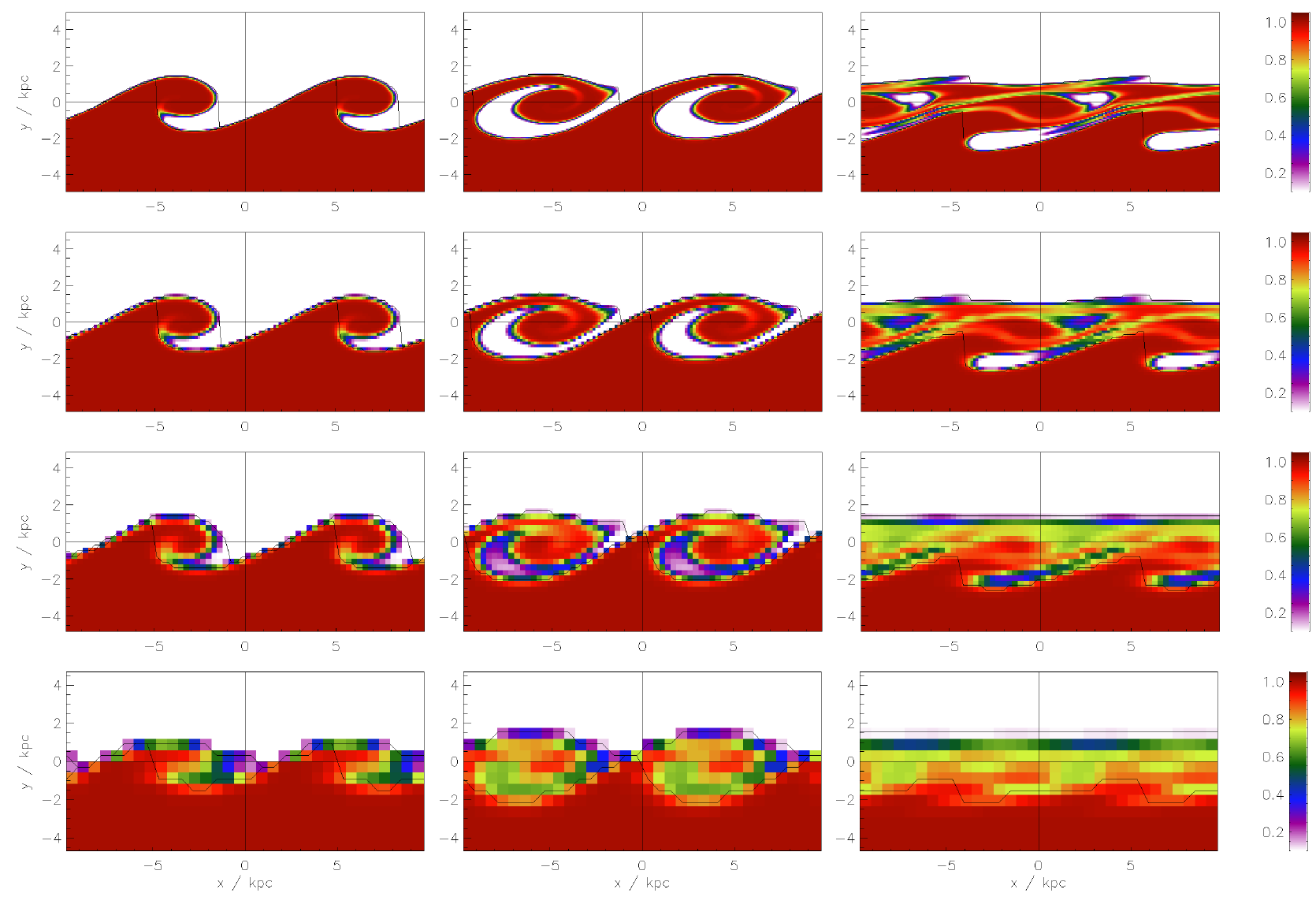}
\caption{
Resolution test for the viscous Kelvin-Helmholtz instability in an idealized setup. The simulation box is periodic in $x$-direction. Pressure is constant throughout the simulation box. Above $x=0$ the fluid temperature is twice as hot as below. The upper fluid moves right,  the lower fluid moves left, their relative velocity is Mach 0.5. Initially the interface is perturbed with a sinusodial vertical velocity $v_y$ of amplitude Mach 0.05 and a wave length of half the simulation box size. We assume a Spitzer-like viscosity with a suppression factor set such that the Reynolds number in the hotter layer is 100. See Roediger et al., in prep., for more details. \newline
The panels show the density of a tracer fluid that was set to 1 in the lower layer and zero in the upper.  Rows from top to bottom are for resolutions of 128, 64, 32, 16 cells per perturbation wavelength. Left, middle and right columns are for timesteps of 50 Myr, 100 Myr, 200 Myr (6.25, 12.5, 24 growth times).  The FLASH code captures the instability even at the lowest resolution.}
\label{fig:extra}
\end{figure*}

Figure~\ref{fig:extra} demonstrates the ability of the FLASH code to capture the KHI even at low resolution.

\end{document}